\documentclass[useAMS,usenatbib]{mn2e}

\usepackage{graphicx}

\bibpunct{(}{)}{;}{a}{}{ }

\begin{document}
\title[NGC\thinspace 253 in colour]{A multi-coloured survey of NGC\thinspace 253 with XMM-Newton: testing the methods used for creating luminosity functions from low-count data}
\author[R. Barnard, L. Shaw Greening and U. Kolb]{R. Barnard$^1$, L. Shaw Greening$^1$ and U. Kolb$^1$\\
$^{1}$Department of Physics and Astronomy, The Open University, Walton Hall, Milton Keynes, MK7 6AA, UK}
\date{}

\pagerange{\pageref{firstpage}--\pageref{lastpage}} \pubyear{2007}

\maketitle

\label{firstpage}

\begin{abstract}
NGC\thinspace 253 is a local, star-bursting spiral galaxy with strong X-ray emission from hot gas, as well as many point sources. We have conducted a spectral survey of the X-ray population of NGC\thinspace 253 using  a deep XMM-Newton observation.    NGC\thinspace 253  only accounts for $\sim$20\% of the XMM-Newton EPIC field of view, allowing us to identify $\sim$100 X-ray sources that are unlikely to be associated with NGC\thinspace 253. Hence we were able to make a direct estimate of contamination from e.g. foreground stars and background galaxies.

 X-ray luminosity functions (XLFs) of galaxy populations are often used to characterise their properties. There are several methods for estimating the luminosities of X-ray sources with few photons. We have obtained spectral fits for the brightest 140 sources in the 2003 XMM-Newton observation of NGC\thinspace 253, and compare the  best fit luminosities of those 69 non-nuclear sources associated with NGC\thinspace 253  with luminosities derived using other methods.
 We find the luminosities obtained from these various methods to vary  systematically by a factor of up to three for the same data; this is largely due to differences in absorption.
 We therefore conclude that assuming Galactic absorption is probably unwise; rather, one should measure the absorption for the population.

 A remarkable  correlation has been reported between the XLFs of galaxies and their star formation rates. However, the XLFs used in that study were obtained using several different methods. If the sample galaxies were revisited and a single method were applied, then this correlation may become stronger still.

 In addition, we find that standard estimations of the  background contribution to the X-ray sources in the field are insufficient. We find that the background AGN may be systematically more luminous than previously expected. However, the excess in our measured AGN XLF with respect to the expected XLF may be due to an as yet unrecognised population associated with NGC\thinspace 253.

\end{abstract}


\section{Introduction}

The X-ray  source populations of external galaxies have been well studied for
the last $\sim$20 years \citep[see e.g.][ and references
  within]{fab89,fab06,rp01,kil05}. Historically, studies of the
individual sources have been severely limited by low count rates and
signal to noise.   One approach is to analyse the colours of each source \citep[see e.g.][]{lmz02, prest03}. Alternatively, one may convert from intensity to flux using an assumed model, and then create an X-ray luminosity function (XLF) to characterise a galaxy or group of galaxies \citep[e.g.][]{fab06,kil05,mis06}.  This model may simply consist of a standard X-ray binary emission model with Galactic line-of-sight absorption \citep[see e.g.][]{zf02, sk02}; alternatively, the model may be obtained from  fitting the whole X-ray population of a galaxy \citep[e.g.][]{irwin03}, or splitting this population into several groups \citep[e.g.][]{rww02}. 
The X-ray point source population for an external galaxy is expected to be dominated by X-ray binaries, with a small fraction being   supernova remnants.

 Empirical relations exist that link the X-ray properties of a galaxy
 with their  mass \citep{gilfanov04} and star formation rate \citep{grimm03}.
However, observations of small or distant galaxies that account for a small portion of the field of view are  dominated by X-ray sources that are unrelated to the target galaxy,  such as foreground stars and background galaxies. Hence, one must estimate the contribution of such sources to the galaxy's XLF before one can estimate its properties. \citet{mor03} have performed one of the most comprehensive studies of the XLFs for the X-ray background to date. They analysed the data from a large number of  deep and wide-field surveys with ROSAT, ASCA, XMM-Newton and Chandra. From these observations they constructed XLFs in a soft band (1--2 keV) and a hard band (2--10 keV), converting from intensity to flux via assumed emission models; these XLFs are normalised by area. Hence, one can estimate the contribution of such background sources to the XLF of a target galaxy by scaling these XLFs by the distance to, and the area covered by, the galaxy.

Using  XMM-Newton, the most sensitive X-ray imaging telescope to date in the 0.3--10 keV band  \citep{turn01,stru01}, we can test the validity of these methods. We can do this by  freely modelling individual X-ray sources in nearby galaxies, and comparing the resulting XLF with XLFs derived using  the various methods described above. NGC\thinspace 253 is ideal for this purpose, as NGC\thinspace 253 is large ($\sim$25$'$$\times$7$'$), but only fills $\sim$20\% of the XMM-Newton field of view. Hence we may study the galaxy and the local background simultaneously. 

NGC\thinspace 253 is a star-bursting spiral  galaxy in the Sculptor group that is almost edge on \citep[inclination  = 78$^{\circ}$, ][]{pen81}. The distance to NGC\thinspace 253 is uncertain, with measurements ranging from 2.58 Mpc \citep{puche91} to 4 Mpc \citep[][]{kar03}. 
The X-ray  population of NGC\thinspace 253 is expected to be dominated by high mass X-ray binaries  (HMXBs), because of the high star formation rate to mass ratio (see e.g. Grimm et al. 2003).

 HMXBs are  classified according to the donor star \citep[see][ and references within for a comprehensive review]{wnp95}. Those with Be star donors have elliptical orbits with periods of  hundreds of days, and only accrete near periastron, via the stellar wind \citep{bh44}; as a result, they are transient sources with luminosities generally $\sim$10$^{33}$--10$^{36}$ erg s$^{-1}$. In HMXBs containing supergiants (SG HMXBs), the donor star is either filling, or almost filling, it's Roche lobe \citep[see ][for a review]{kap04}. The compact object in SG HMXBs is continuously accreting either via Bondi-Hoyle accretion  or via an accretion disc; Bondi-Hoyle accretion yields luminosities in the range  $\sim$10$^{33}$--10$^{36}$ erg s$^{-1}$, while disc accretion can power luminosities up to $\sim$10$^{38}$ erg s$^{-1}$. Our survey is limited to bright X-ray sources, and hence we expect our sample to be dominated by disc-fed sources.

 In this work we examine the  2003, June XMM-Newton observation of NGC\thinspace 253. We first provide details of the observation and the data analysis in Sect. 2.  We then present our results in Sect. 3. We first freely model the spectra of 140 bright X-ray sources, including  69 that we associate with NGC\thinspace 253, and examine their group properties. We then   compare the luminosities obtained from  these spectral fits with luminosities obtained using several standard methods. Finally we compare the theoretical AGN XLF of \citet{mor03} with the XLF derived from our freely fitted background sources.  We discuss the implications of our findings in Sect. 4, particularly the implications for the empirical relation between the X-ray properties of a galaxy and its star formation rate,  and the universal HMXB XLF proposed by \citet{grimm03}. Finally, we draw our conclusions in Sect. 5. 
\begin{figure}
\resizebox{\hsize}{!}{\includegraphics[angle=0,scale=0.6]{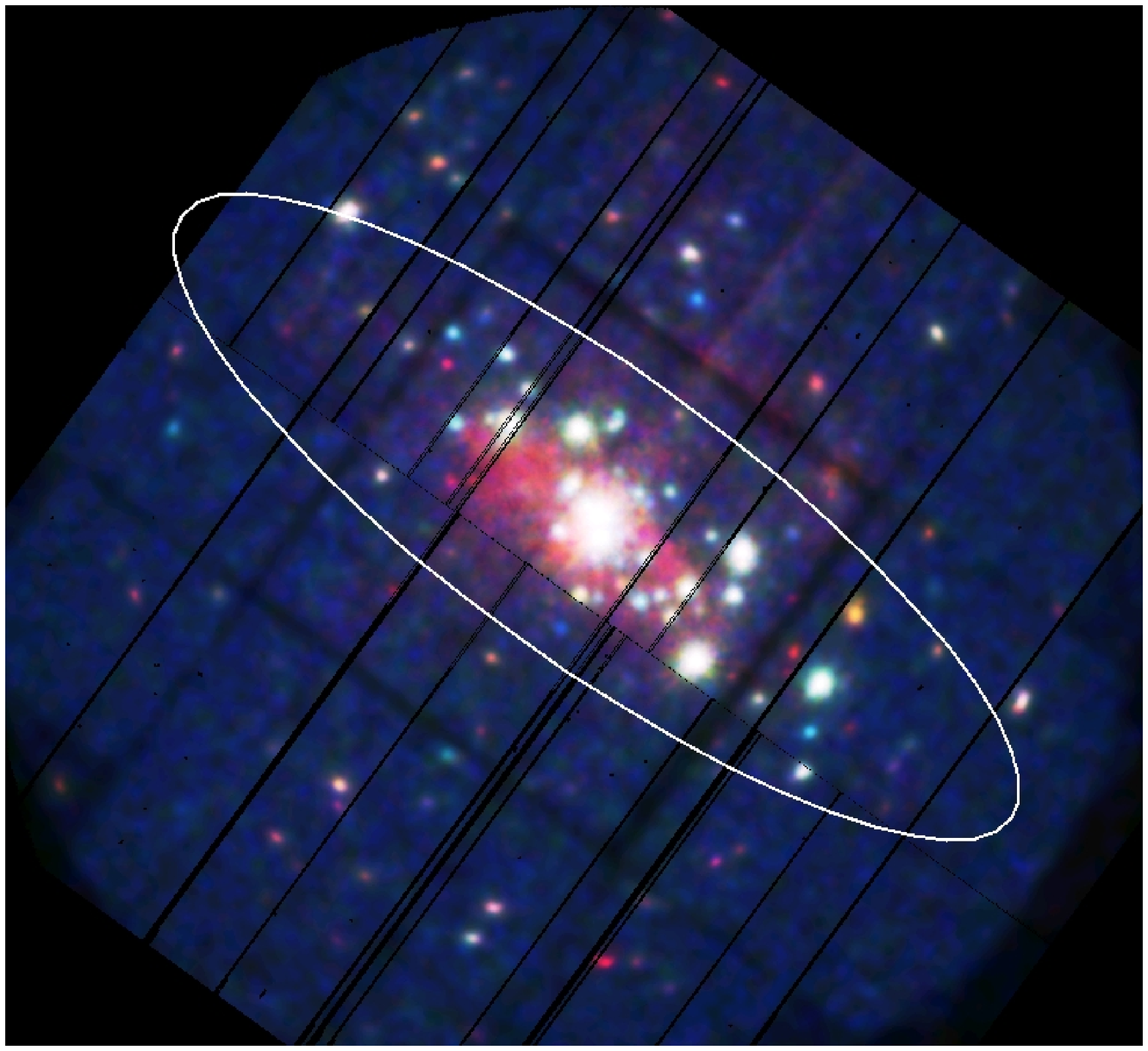}}
\resizebox{\hsize}{!}{\includegraphics[angle=0,scale=0.6]{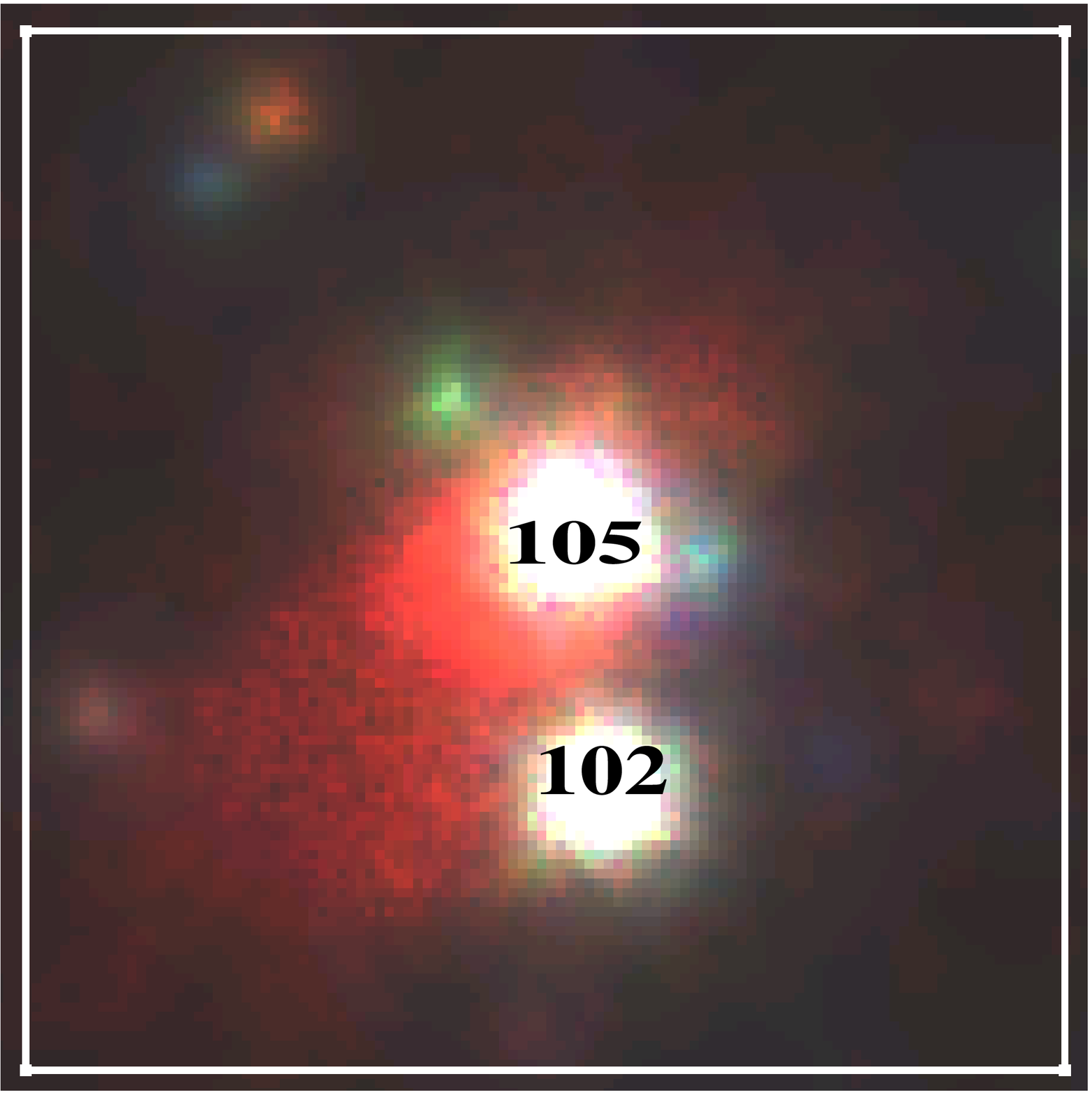}}
\caption{ {\em Upper panel}: Three-colour, combined EPIC, $\sim$30$'$$\times$ 30$'$ image from the 2003 XMM-Newton observation of NGC\thinspace 253; the intensity scale is histogram equalised. North is up, East is left. The energy bands used were 0.3--2.0 keV (red), 2.0--4.0 keV (green) and 4.0--10.0 keV (blue). The white ellipse represents the V band  D$_{25}$ contour of NGC\thinspace 253. {\em Lower panel}: Linear-scaled image of the central 2$'\times 2'$ region; several  point sources are visible, while the red smudge is  the superwind reported by Pietsch et al. (2001). Source 105 is the nuclear region.}\label{3im} 
\end{figure}

\section{Observations and data analysis}

XMM-Newton observations are susceptible to periods of high background levels, caused by increased flux of solar particles. We screened the data from each of the EPIC cameras (MOS1, MOS2 and pn), to remove flaring intervals.
   This process resulted in $\sim$46 ks of good time for the pn and $\sim$69 ks for the MOS cameras.

We combined the cleaned MOS and pn  data, and ran the  source detection algorithm provided with the XMM-Newton data analysis suite SAS version 7.0. We accepted sources with maximum likelihood detections $>$10 (equivalent to 4$\sigma$). For every source, we obtained an extraction region
with radius 12--40$''$. In general, we used a radius of 20$''$, 
  except for sources with large PSFs due to high off-axis angle, where
  a 40$''$ radius was used, or in very crowded regions, where a
radius of 12--15$''$ was used. The extraction radius for each
  source is provided in Table A1. The central region of NGC\thinspace 253 is
  fairly crowded, and we ignored sources with another source $<$10$''$
  away. This resulted in the loss of only a few faint sources. 

We also created a corresponding background region for every source. We required that the background be on the same CCD as the source for all three EPIC cameras, that there be no point sources in the background, and that its intensity per unit area be smaller than for the source region. The resulting background regions had areas 1--35 times greater than their corresponding source regions;  for 75\% of sources, the background area was more than three times larger than the source area.

We extracted pn and MOS  source and background spectra in the 0.3--10 keV range, combining the MOS1 and MOS2 spectra if the source was present in both cameras.
We obtained fits to all spectra with $>$50 source counts in  the pn
 and/or MOS spectra   with XSPEC 11.3\footnote[1]{http://heasarc.nasa.gov/xanadu/xspec/}. We used power law, blackbody and
bremsstrahlung models; all models included line-of-sight
absorption. We considered all fits with null hypothesis probability
$\geq$0.05 as acceptable; this is the probability that the
  differences between the  modelled and observed spectra are due to
  random fluctuations alone. If none of these spectral models provided a good fit, we considered a two-component model consisting of a power law and blackbody, as seen in Galactic X-ray binaries. We used the best fit model to obtain a 0.3-10 keV, unabsorbed flux for each source.

\section{Results}

In Fig.~\ref{3im} we present a three-colour, combined EPIC image
($\sim 30'\times 30'$) from  the 2003, June observation. The images were binned to
2400$\times$2400 pixels,  then smoothed with the SAS task {\em
  asmooth}. The smoothed images were weighted by corresponding
exposure maps.  North is up, East is left. The white ellipse
represents the V band D$_{25}$ isophot for NGC\thinspace 253.  Both populations of X-ray sources, inside and outside the
D$_{25}$ isophot, clearly display a wide range of colours.  We
  also show a linearly-scaled close-up of the central 2$'\times 2'$
  region, showing several point sources in a region that looks like
  one unresolved source in the main image.

\begin{table*}
 \centering
 \caption{Best fit parameters for power law models applied to the summed spectra of the IS and OS faint source populations.  We first show the number of faint sources in the population.  We  then show the best fit absorption and photon index, the $\chi^2$/dof and corresponding good fit probability, and the flux equivalent to 1 count s$^{-1}$ in the 0.03--10 keV band for an extraction region with 15$''$ radius in the pn and MOS.  Numbers in parentheses indicate 90\% uncertainties in the last digit}\label{specs}
  \begin{tabular}{cccccccc}
  \noalign{\smallskip}
  \hline
  \noalign{\smallskip}
   Model  & $N_{\rm Fnt}$ & $n_{\rm H}$ / 10$^{20}$   & $\Gamma$ & $\chi^2$/dof &  Flux (pn thin) & Flux (MOS med)\\
\noalign{\smallskip}
 \hline
\noalign{\smallskip}  
IS& 7 & 1.3 & 0.4(3) & 71/54 [0.06] & 1.1(2)E+4 & 3.6(8)E+4 \\
OS& 36 & 1.3 & 1.23(13) & 55/44[0.12] & 4800(400) & 1.9(2)E+4\\
\noalign{\smallskip}
\hline
\noalign{\smallskip}
\end{tabular}
\end{table*}

\begin{figure}
\resizebox{\hsize}{!}{\includegraphics[angle=270,scale=0.6]{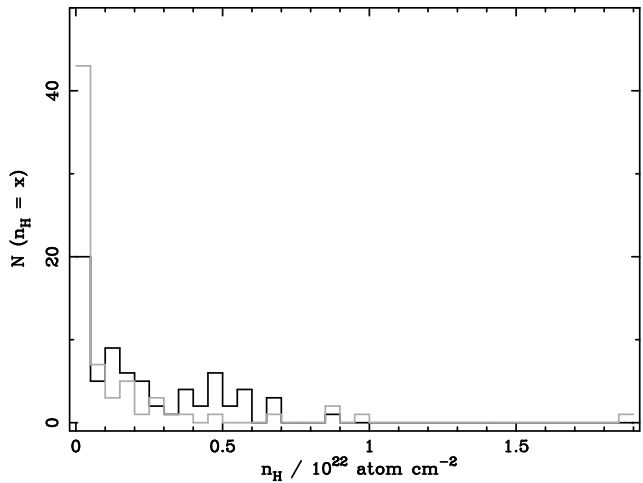}}
\caption{ Histograms of best fit  line-of-sight absorption, $n_{\rm H}$, for the IS (black) and OS (grey) populations.  }\label{nhhist2} 
\end{figure}

\begin{figure}
\resizebox{\hsize}{!}{\includegraphics[angle=270,scale=0.6]{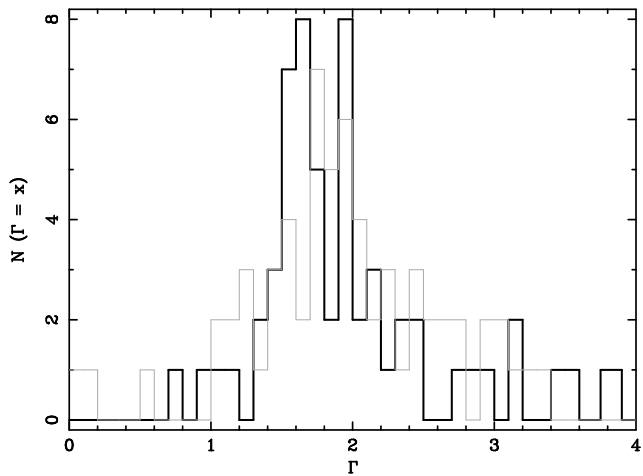}}
\caption{Histograms of best fit spectral index, $\Gamma$,  for the IS (black) and OS (grey) populations.}\label{phist2} 
\end{figure}

\subsection{Point source detection}

Source detection revealed 185 point sources;    they are designated
S1--S185, and their locations are presented in Table~\ref{pos} in the
Appendix. In this work we separate these sources into  those inside
and outside the D$_{25}$ isophot of NGC\thinspace 253,  IS and OS
respectively.  We found  140 out
of the 185 XMM-Newton sources  in Chandra observations: 3 XMM-Newton detections contained multiple Chandra sources, but the other 137 are single point sources, as far as can be told by the current X-ray telescopes. 
\citet{vp99} identified S9, S65 and S163 as background QSOs via their optical counterparts; they also identified S100 as a foreground star. \citet{vp99} also labelled S102 as a possible black hole X-ray binary, due to its high luminosity.

 We found no associations between our X-ray sources and the identified globular clusters;   however, we found two X-ray sources that were within 1$''$ of globular cluster candidates \citep{bs00}. One of these was S65,  leaving just S166 as a possible globular cluster X-ray source.  These results suggest that the population outside the D$_{\rm 25}$ region are almost entirely unrelated to NGC\thinspace 253, making it a good probe of the local X-ray background. However, we note that \citet{gal04} have announced 380 globular cluster candidates in NGC\thinspace 253, but have yet to publish the results of their follow-up observations.

\subsection{Spectral analysis}

 We found 71 IS and 69 OS sources   to be  bright enough for spectral modelling.  Lightcurves of each source were checked for variability; the spectral models are only valid if the source is stable. Variability of sources in NGC\thinspace 253 will be discussed in a following paper (Barnard et al. in prep).

Our strategy for obtaining spectra was designed to ensure at least 5-10 source counts per channel, favouring a few 10-count channels over many channels with less source counts. Spectra with $>$500 source counts were grouped to a minimum of 50 counts per bin; those with 200--499 were grouped to a minimum of 20 counts per bin; those with 50--199 counts where the source contributed $\ge$50\% of the total counts were binned to a minimum of 10 counts per bin; finally, sources with 50--199 counts where the source contribution $<50$\% were grouped to a minimum of 20 counts per bin. We also note that the source + background spectrum  always had at least 10 counts per channel by design, and that our large background regions  ensured a good determination of the background spectrum. As a result, even sources with only 51 source counts could discriminate between models in some cases.

\subsubsection{Spectral properties of the  bright X-ray sources}

Here we compare the spectral properties of those sources bright enough for modelling.  Details of each fit are presented in Table A2. We first looked at the range in line-of-sight absorption exhibited by these sources, then looked at the best fit photon index for a power law model, even for sources where a power law does not give the best fit.  For this comparison we ignored S105 (the nuclear region) and S163 (an AGN) from the IS population, and S100 (a foreground star) from the OS population.

  Figure~\ref{nhhist2} shows  histograms  of the line-of-sight
  absorption, $n_{\rm H}$, for the IS (thick black) and OS (thin grey)
  populations, with a resolution of 0.05$\times$10$^{22}$ H atom cm$^{-2}$.  We see that $\sim$70\%
  of the IS population have absorptions $>$5$\times$10$^{20}$ H atom
  cm$^{-2}$, i.e. $>$4 times Galactic line-of-sight absorption   \citep[$\sim$1.3$\times 10^{20}$ H atom cm$^{-2}$, ][]{stark92};
  indeed, 16 of the sources ($\sim 25\%$) exhibit absorption $>$40
  times Galactic absorption. This variation is perhaps  unsurprising
  in a spiral galaxy that is almost edge on, particularly when one
  expects the X-ray sources in NGC\thinspace 253 to be HMXBs, and therefore
  linked to regions of high star formation rate. Meanwhile, the
  majority of the spectrally fitted OS sample, thought mostly to be
  background galaxies,  have low absorptions. This is likely to be a
  selection effect, caused by choosing only the brightest
  (i.e. nearest, or least absorbed) galaxies.

Figure~\ref{phist2} shows the distribution in  photon index, $\Gamma$,
for the best fit power law models to the  IS and OS populations.  Again, the IS histogram is represented by a thick black
line  and the OS histogram by a thin grey line.  Around 10\% of sources
exhibited thermal spectra, and yielded $\Gamma$ $>$ 4; such sources
were  excluded from Fig.~\ref{phist2}. While  many
published XLFs assume a single value of $\Gamma$, a broad range is observed in both the IS and OS populations.

\subsubsection{Modelling the faint sources}

 We classify those sources  that are not bright enough for spectral modelling as faint sources.
We separately summed the spectra of the  7 faint  IS and 36 faint OS sources, and modelled each of the  summed spectra  with a best fit power law.  Table~\ref{specs} summarises these models. For each model, we give the best fit absorption and photon index, as well as the conversion from intensity to flux for each instrument in the 0.3--10 keV band.  This conversion factor is defined as the unabsorbed flux equivalent to 1 count s$^{-1}$ for an on-axis source with an extraction radius of 15$''$. We initially used the HEASARC WebPIMMS software\footnote[2]{http://heasarc.gsfc.nasa.gov/Tools/w3pimms.html} to calculate the conversion factors for each model. However, we realised that WebPIMMS does not account for the variation in calibration throughout the lifetime of XMM-Newton. Hence, we obtained the conversion factors for the observation discussed here using an on-axis, 15$''$ source. 
 We note that $\Gamma$ = 0.4$\pm$0.3 for the best fit IS power law model. This is consistent with the 0.3--10 keV spectra of  faint NS+Be HMXBs accreting via Bondi-Hoyle accretion \citep[where $\Gamma$ $\sim$0--1.5, ][]{wnp95}.

\begin{table*}
 \centering
 \caption{ Best fit absorbed power law models used to obtain the conversion factors for Methods I--III; the conversion factor is the 0.3--10 keV unabsorbed flux  equivalent to a 0.3--10 keV pn intensity of  1 count s$^{-1}$ from an on-axis source region with 15$''$ radius. For each model we give $n_{\rm h}$, $\Gamma$, $\chi^2$/dof and conversion factor (0.3--10 keV unabsorbed flux in 10$^{-15}$ erg cm$^{-2}$ s$^{-1}$ equivalent to 1 count s$^{-1}$ in the pn  from an on-axis source with 15$''$ radius). }\label{cf}
  \begin{tabular}{rccccccc}

\noalign{\smallskip}
\hline
  \noalign{\smallskip}
Model & $n_{\rm H}$ / 10$^{20}$ atom cm$^{-2}$ & $\Gamma$ & $\chi^2$/dof & Conversion Factor\\
\noalign{\smallskip}
\hline
\noalign{\smallskip}
 Model I: Standard model  & 1.3 & 1.7 & 9909/779 & 4006\\
Model II: Best fit to all non-nuclear NGC\thinspace 253 sources  & 34 & 2.49 & 1339/777& 8252 \\
Model III: $<$3000 counts  & 19 & 2.40 &665/495 & 5365\\
Model III: $>$3000 counts & 50 & 2.65 & 1795/1959  &11693\\
 \noalign{\smallskip}
\hline
\noalign{\smallskip}
\end{tabular}
\end{table*}

\begin{figure*}
\resizebox{\hsize}{!}{\includegraphics[angle=270,scale=0.6]{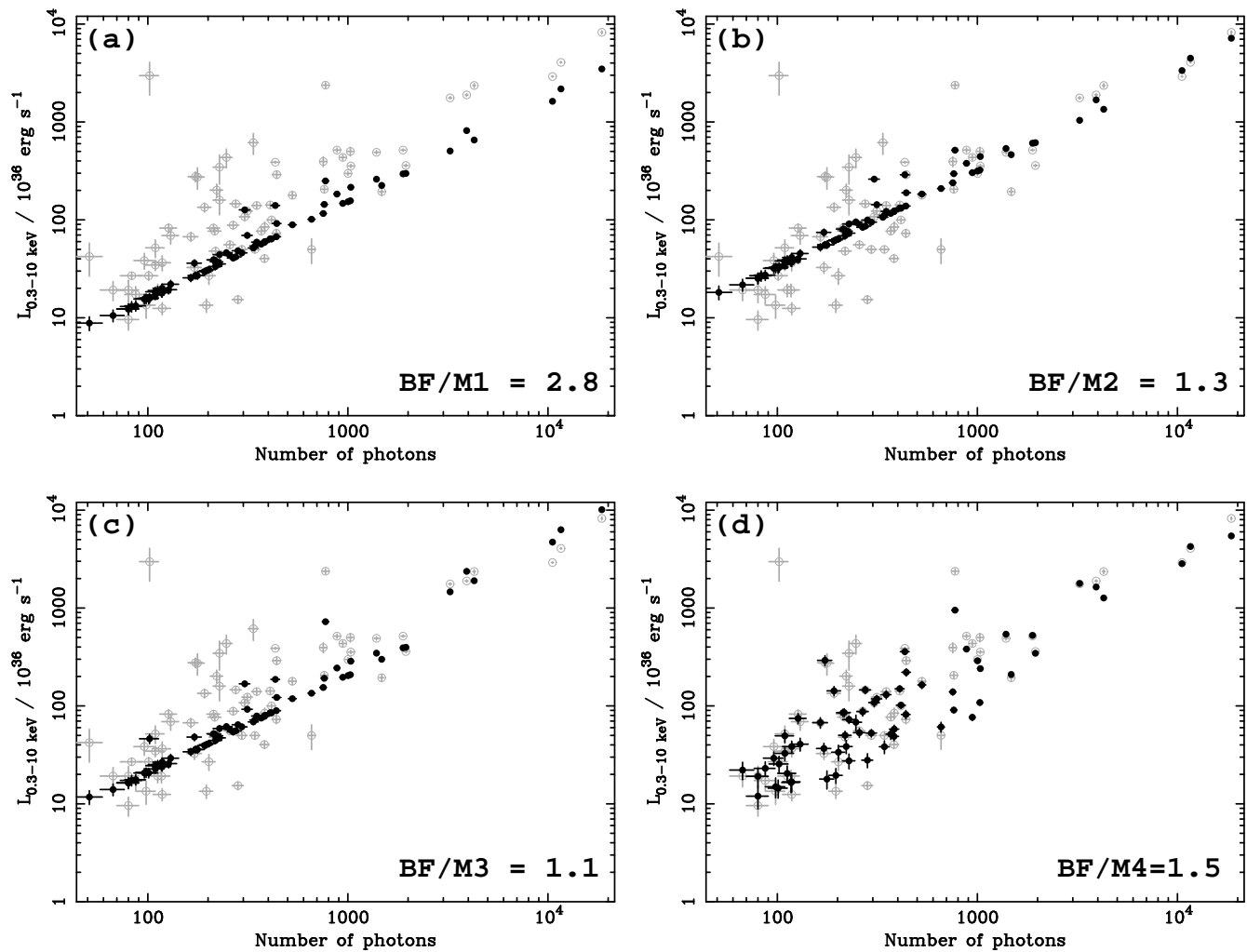}}
\caption{  Comparison of luminosities obtained from best fits to  69 IS sources  with luminosities obtained using Methods I--IV
  (panels a--d respectively) as a function of the number of pn source
  photons accumulated from the source. A distance of 4 Mpc is assumed. The best fit luminosities are grey, hollow circles, while luminosities derived using each method are dark, solid circles. For each model we give the ratio BF/M$x$, where $x$ is 1--4. This is the ratio of the summed 0.3--10 keV luminosity from the individual best fits to the summed 0.3--10 keV luminosity using Method I--IV.}\label{bfcomp} 
\end{figure*}

\subsection{Luminosities of NGC\thinspace 253 X-ray sources from different methods}

\subsubsection{ Defining the methods}
For the 69  IS sources discussed in Sect. 3.2.1, we obtained fluxes from the source intensities using some of the methods employed  in the literature  when creating the XLFs of external galaxies. 

 For Method I, we assumed a standard emission model and Galactic line-of-sight absorption. We tried a 5 keV bremsstrahlung model \citep[see e.g.][]{zf02} as well as a power law with $\Gamma$ = 1.7 \citep[see e.g.][]{sk02}; we found the conversion factors for these models to agree within 3\% for the pn and 5\% for the MOS.  We  chose the power law  model. For Method II, we combined the source regions for  all of the  sources, and obtained a  composite  source spectrum; similarly, a  composite  background spectrum was obtained by adding all the background regions. The background-subtracted spectrum was then modelled by an absorbed power law, giving the conversion factor; this method is similar to that  used by e.g. \citet{irwin03}.  Method III involved splitting the NGC\thinspace 253 population into sources with $>$3000 net counts  and those with $<$3000 net counts. We then found the conversion factor for each group. This method is similar to that employed by e.g. \citet{rww02}. Finally, we  modelled each source  with constrained power law emission with $\Gamma$ = 1.7, but with the absorption free to vary; the motivation was to investigate the importance of the emission spectrum in deriving the source luminosity.  We call this Method IV.
 Luminosities were calculated assuming a distance of 4 Mpc, as favoured by \citet{grimm03}.
 
\subsubsection{Applying the conversion factors}
 In Table~\ref{cf}, we list the spectral models and conversion factors used for Methods I-III. For each model, we give $n_{\rm H}$, $\Gamma$, and  $\chi^2$/dof, as well as the conversion factor.
  This conversion from source intensity to flux assumes an on-axis source with a 15$''$ radius; however, our source regions varied in radius from 12$''$--40$''$, and had off-axis angles of $\sim$0--14$'$.  Hence, it was necessary to correct the source intensities for vignetting, and differences in encircled energy fraction (EEF). 
The background-subtracted source intensities obtained from XSPEC are already vignetting corrected, and only EEF correction was necessary. We calculated the EEF for every source as a function of radius and off-axis angle, weighted by energy over the 0.3--10 keV band. If $I( R, \Theta)$ and $E(R, \Theta)$ are respectively the intensity and EEF for a source region of radius $R$  and off-axis angle $\Theta$, then
\begin{equation}
\frac{I(R,\Theta)}{E(R,\Theta)} = \frac{I(15,0)}{E(15,0)},
\end{equation}
hence 
\begin{equation}
I(15,0) = I(R,\Theta)\times \frac{E(15,0)}{E(R,\Theta)}.
\end{equation}  
 This work ensures that the same corrections for EEF and vignetting were applied for Methods I--IV as for the best fit spectra. We present the EEF for each source in Table A1. $E(15,0)$ was found to be 0.71 for the pn,  0.68 for MOS1 and 0.69 for MOS2.

\subsubsection{Comparing the methods}

In Fig.~\ref{bfcomp} we compare the best fit luminosities with
luminosities derived from Methods I--IV as a function of pn source
counts in panels
(a)--(d) respectively. The best fit luminosities are shown as grey,
open circles, while the luminosities derived from Methods I--IV are
solid circles.  For Methods I and II, the luminosities are expected to
have a linear relation to the number of source counts, and the scatter is  due to differences in encircled energy and vignetting corrections.

 We also show the ratio of total  best fit to modelled luminosity for
 each method (BF/M$x$, where $x$ is the method number from 1 to 4).
 The total luminosity  from  the best fit spectra is 2.8 times higher than the Model I luminosity  for the same sources,  30\% higher than for Model II, and 10\% higher than for Model III. 
It is unsurprising that Model III is most successful at reproducing
the freely fit luminosity, as  we grouped the sources into two intensity groups and obtained  two very different conversion factors. Hence the relationship between intensity and luminosity is clearly non-linear.

We note with interest that Model I, with a best fit $\chi^2$/dof $\sim$13, is furthest from agreement with individual fits. Meanwhile, the best fit   for the Model III sources with  $>$3000 counts yields $\chi^2$/dof $\sim$0.92, and BF/M3 for  just those sources is just 1.03. These results show that reliable luminosities can only be obtained from models that reflect the data. 

For low luminosity sources, the Method IV and best fit luminosities agree fairly well. This is to be expected, as many low-luminosity disc-accreting XBs exhibit spectra that are  well characterised by $\Gamma$=1.7. However, at higher luminosities, Method IV tends to underestimate the luminosity. Again, this is expected, as higher luminosity XBs exhibit systematically softer spectra than low luminosity XBs; hence $\Gamma$=1.7 is generally no longer a good fit. As a result, Method III and even Method II give a better estimate of the integrated lumiosity of NGC\thinspace 253.

\subsection{Luminosity functions of the NGC\thinspace 253  and background populations}
\label{lf}

\begin{figure}
\resizebox{\hsize}{!}{\includegraphics[angle=270,scale=0.6]{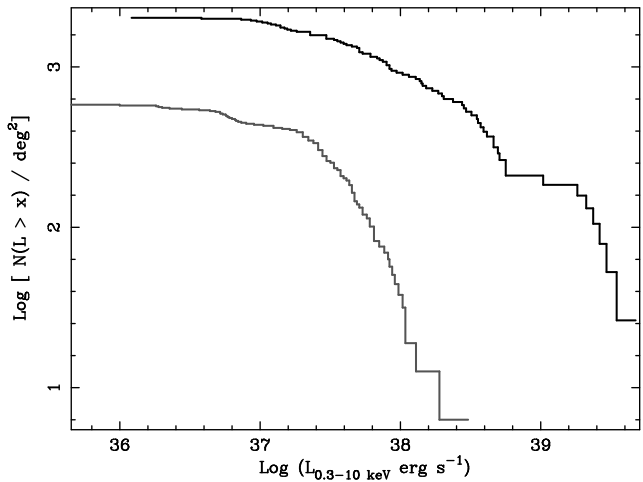}}
\caption{ Best fit X-ray luminosity functions (XLFs) for the IS (black) and OS (grey) populations, assuming a distance of 4 Mpc. The nuclear source S105 and QSOs S163 and S26 are removed from the IS XLF. The foreground star S100 is removed from the OS XLF because it suffers 100\% uncertainties. Kolmogorov-Smirnov (K-S) testing shows that the IS and OS populations have a probability of 1.8$\times$10$^{-7}$ for being drawn from the same population.}\label{lf1_2} 
\end{figure}

 While we associate the IS population with NGC\thinspace 253, the OS population represents the background AGN.  Before comparing the XLFs of the IS and OS populations, we normalised them by area. Assuming a circular field of view with 15$'$ radius,  NGC\thinspace 253  and  the background region cover 137 and 570 square arcminutes respectively. We present the best-fit XLFs of the IS and OS populations   in Fig.~\ref{lf1_2}; the 0.3--10 keV  luminosity is plotted on the x-axis, and the number of sources per square degree  with higher fluxes given on the y-axis.   The black lines represent the IS XLF, while the grey lines represent the OS XLF.  These luminosities are calculated assuming  a distance of 4 Mpc. It is clear that the IS population has considerably higher spatial density than the OS population. Figure~\ref{lf1_2} leads us to  expect little contribution from the background  above $\sim$2$\times$10$^{37}$ erg s$^{-1}$.

\begin{figure}
\resizebox{\hsize}{!}{\includegraphics[angle=270,scale=0.6]{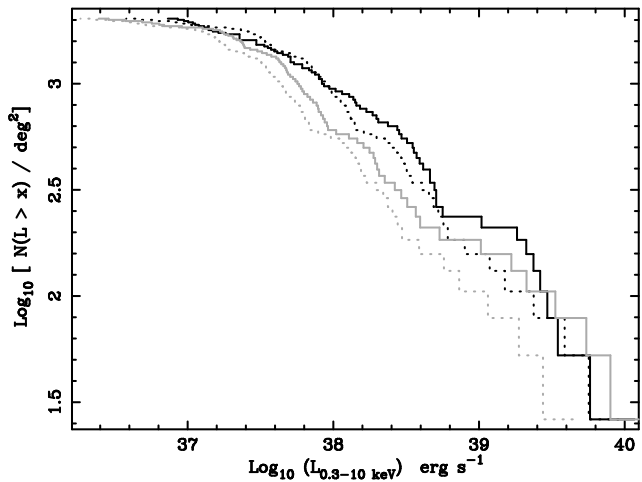}}
\caption{ Comparison of best fit NGC\thinspace 253 XLF (black, solid) with XLFs
  obtained using  Method I (grey dotted), Method II (black, dotted) and
  Method III (grey, solid). }\label{lfcomp} 
\end{figure}

We next compare the best fit IS XLF with those of Methods I--III in
Fig.~\ref{lfcomp}. Unsurprisingly, the XLFs of Methods II and III are
flatter than the XLF of Method I, as the Method I luminosities are
systematically lower than for Methods II and III. The universal XLF for HMXBs reported by \citet{grimm03} relied on published XLFs that were created using a variety of methods. Our results show that the systematic variations between methods is likely to have introduced extra uncertainties in their results; we discuss the implications of our results more fully in Sect.~\ref{ulxf}.

\begin{figure}
\resizebox{\hsize}{!}{\includegraphics[angle=270,scale=0.6]{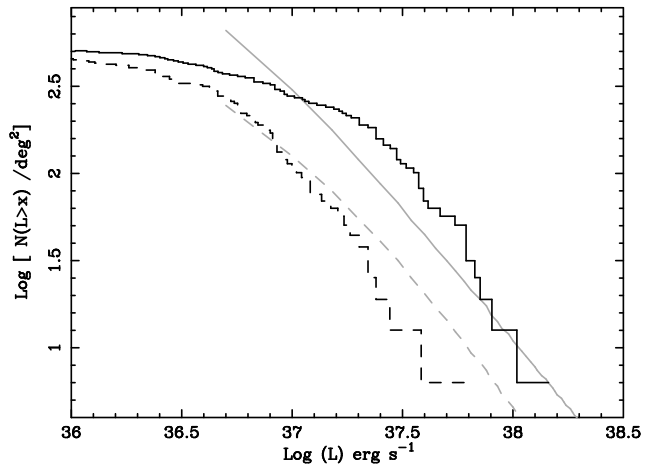}}
\caption{  Comparison of the OS XLF (black) with the \citet{mor03} XLFs  (grey); 1--2 keV XLFs are represented by dotted lines, while the 2--10 keV LFs are solid.  The OS XLF includes sources that were too faint to fit individually, so the best fit power law to the summed faint source spectrum was used (see Table~\ref{specs}). The \citet{mor03} XLFs were constructed from several surveys where fluxes were obtained by assuming a single spectral model for a given field.  Our XLFs suggest that the true contribution of background AGN is higher in the 2--10 keV band, and somewhat lower in the 1--2 keV band, than is suggested by \citet{mor03}.}\label{lf2_m} 
\end{figure}

\subsection{Assessing the AGN contribution}

 In Fig.~\ref{lf2_m} we compare the \citet{mor03} XLFs with best fit OS XLFs in the 1--2 keV (soft) and 2--10 keV (hard) bands, including faint sources. For the faint sources, the OS model presented in Table 1  is used to convert from intensity to flux.    We note that our OS sample is rather limited, and does not necessarily represent the AGN background as a whole. However, our best fit OS XLF suggests that AGNs contribute rather more to the high luminosity end of the hard XLF than predicted by \citet{mor03}; conversely, our background AGNs appear to contribute less to the soft XLF than predicted. We also note that the best fit OS XLFs are steeper than those calculated by \citet{mor03}.  Our observed 2--10 keV XLF  for luminosities greater than $\sim$10$^{37}$ erg s$^{-1}$ suggests that the AGN are systematically more luminous than previously supposed; incompleteness prevents us from exploring the low luminosity end of the XLF. The observed excess could be due to an as yet unidentified population associated with NGC\thinspace 253. However the lack of globular cluster associations to date makes this unlikely. Alternatively, the spectra of the AGN could simply be harder than assumed by \citet{mor03}, who assign an emission model to each source based on its intensity.

\section{Discussion}

\subsection{Comparing Methods I--IV}

Absorption appears to be a major influence on the conversion factors for Models I--III;  the modelled absorptions for Methods II and III are $\sim$15--40 times higher than the Galactic line-of-sight absorption assumed for Method I. As a result, the luminosities obtained from Methods I--III vary by a factor of $\sim$3. However, our results from Method IV show  that absorption is not the only important influence in determining the source luminosity.   We therefore conclude that one should not assume the absorption or emission spectra of extragalactic X-ray sources.

We expect many of the X-ray sources in NGC\thinspace 253 to be disc-fed HMXBs. Galactic disc-fed LMXBs and HMXBs are known to exhibit softer spectra at higher luminosities, whether the accretor is a neutron star or black hole \citep[see e.g.][and references within]{vdk94,vdk95,mr03}. Hence it is unsurprising that the relationship between counts and luminosity is non-linear. Subdivision of the X-ray population into intensity bands, and obtaining corresponding conversion factors (like Roberts et al., 2002), is therefore likely to be the best approach to fitting any low-photon-count  data. 

We note that \citet{irwin03} used an approach very similar to  Method II for studying LMXBs in Chandra observations in nearby elliptical galaxies. They modelled the composite spectrum for each galaxy with an absorbed power law, but fixed the absorption in each case to  Galactic values.  They obtained good fits for each galaxy, with $\Gamma$ ranging over $\sim$1.4--1.9 between galaxies. These results suggest that using Galactic absorption and $\Gamma$ = 1.7 (i.e. Method I) is acceptable for  LMXBs in elliptical galaxies. However, we would still recommend Method II (preferably Method III), in case the X-ray population to be studied experiences absorption significantly higher than Galactic line-of-sight, or is systematically harder or softer than expected.

\subsection{ Implications for the universal HMXB  XLF }
\label{ulxf}

\citet{grimm03} used Chandra and ASCA surveys of nearby starburst galaxies, along with ASCA, MIR-KVANT/TTM and RXTE/ASM observations of HMXBs in our Galaxy and the Magellanic Clouds to obtain a correlation between the X-ray properties of HMXB populations and the star formation rate (SFR) of their host galaxies. They chose their sample of galaxies to have sufficiently high SFR to total mass ratios so that their X-ray populations would be dominated by HMXBs, with negligible LMXB contributions.

\citet{grimm03} used published Chandra XLFs, scaled to distances calculated from the radial velocities of \citet{st80}, assuming the Hubble constant to be 70 km s$^{-1}$ Mpc$^{-1}$. They found the XLFs of these galaxies to be strikingly similar, when normalised by the SFR of the galaxy; estimates for the SFR were obtained from IR, UV, H$_{\alpha}$ and radio observations.

 \citet{grimm03} proposed a universal HMXB XLF, with the differential form
\begin{equation}
\frac{{\rm d}N}{{\rm d}L_{38}} = \left(3.3^{+1.1}_{-0.8}\right)\times {\rm SFR}\times  L_{38}^{-1.61\pm0.02},
\end{equation}
where $L_{38}$ is the luminosity normalised to 10$^{38}$ erg s$^{-1}$, and the SFR is measured in M$_{\odot}$ yr$^{-1}$. They also derived two empirical relations between the X-ray properties of their sample galaxies and the SFR. Firstly, they find that the number of sources with 2--10 keV luminosities $>$2$\times$10$^{38}$ erg s$^{-1}$ to be proportional to SFR$^{1.06\pm0.07}$. Secondly, they find a linear relation between the total HMXB X-ray flux of a galaxy and its SFR, for SFRs $\ga$4 M$_{\odot}$ yr$^{-1}$; at lower SFRs, they find the total luminosity to be proportional to SFR$^{1.7}$.

\citet{grimm03} included NGC\thinspace 253 in a secondary sample of galaxies used to test their XLF-SFR relation. They obtained the integrated 2--10 keV luminosity of 5$\times$10$^{39}$ erg s$^{-1}$ from an RXTE observation, using the best fit to the integrated spectrum (a power law with $\Gamma$ = 2.7), and assuming a distance of 4 Mpc.  The SFR estimates from the multi-wavelength observations  varied from 1.5 M$_{\odot}$ yr$^{-1}$ (H$_{\alpha}$) to  6.5--9.5  M$_{\odot}$ yr$^{-1}$  (FIR); \citet{grimm03} adopted a SFR of 4 M$_{\odot}$  yr$^{-1}$. Using their values, NGC\thinspace 253 is  considerably less luminous than predicted by their relations.

However, the XLFs used by \citet{grimm03} were created using several
methods. Some were obtained assuming a standard emission model and
Galactic absorption \citep[e.g.][]{zf02,sk02}, while \citet{kar01}  used a single emission model derived from source
colours. \citet{esc02} assumed Galactic absorption for their XLFs,
but derived spectra of each source from their colour and intensity,
while \citet{lmz02} derived the absorption and emission
characteristics of their sources from the source colours. \citet{sw01}
derived their XLF from the best fit spectral models to the brightest
X-ray sources, while \citet{rww02} grouped their sources by intensity
and derived conversion factors for each group. 

 \citet{grimm03} have shown a striking correlation between the star formation rates and XLFs of different galaxies, although there is sigificant scatter in the correlation.
However, our results have shown that one can obtain very different luminosities
from the same data when using different methods. Hence the differences in the XLFs presented by \citet{grimm03} may be due to the different methods used in obtaining each XLF. If one were to revisit the galaxies surveyed by \citet{grimm03} and obtain the XLFs using only a single method (preferably Method III), then the correlation between XLF and star formation rate may be strengthened.

\section{Conclusions}

Grimm et al. (2003) report a universal HMXB XLF derived from published XLFs of several nearby galaxies. They also derive relations between the star formation rate and (i) the total luminosity of the point X-ray sources in the galaxies and (ii) the number of X-ray sources in a galaxy with 2--10 keV luminosity $>$2$\times$10$^{38}$ erg s$^{-1}$.   However, the published XLFs were produced using a number of methods, in most cases assuming Galactic line-of-sight absorption. 

We have tested several of these models using a deep XMM-Newton observation of the nearby galaxy NGC\thinspace 253, included in a secondary sample of Grimm et al. (2003). We obtained freely modelled luminosities for the 140 brightest sources in the field and also obtained the conversion factors from intensity to flux for some of these different models. We  found them to vary by a factor of $\sim$3.   We found the biggest influence on the conversion factor to be the absorption, which varied by a factor of $\sim$50 between methods. Since the universal XLF and relations between SFR and X-ray properties were obtained using a mixture of methods,  we suggest that reanalysing the sample with a single approach could yield  even more striking relations. 

It is possible that NGC\thinspace 253 may represent a ``worst case'' for absorption effects, as it is almost edge on and has a high star formation rate. However we note that our early (unpublished) studies of XMM-Newton observations of NGC\thinspace 300 show similar disparity between Method I and best fit luminosities; NGC\thinspace 300 is almost face on and has a much lower star formation rate. Hence, our concerns about assuming Galactic absorption and standard emission models may apply to all galaxies to some extent.

Furthermore, we find that the background XLFs obtained by \citet{hasinger01} and \citet{mor03} should be treated with caution and may  be  misleading. The true background XLF appears to contribute more to the high-flux sources, and correspondingly less to the lower fluxes. I.e., the high flux gradient is steeper, while the low flux gradient is  flatter. Hence, the background XLFs also need to be re-calibrated using best fit models rather than assumed models. 

 It is not yet possible to combine the superlative spatial resolution of Chandra with the sensitivity of XMM, hence recalibrating the XLF-SFR relation, and the XLF for background AGN,  will be difficult. However, we find that using an intensity to flux conversion that is derived from the best fit model to summed X-ray sources is more successful than just assuming a standard model. The more luminous X-ray sources are systematically softer than the fainter ones, hence it makes sense to group X-ray sources by intensity, and apply the best fit to the summed spectrum of each group when converting from intensity to flux.

\section*{Acknowledgments}

 We gratefully thank the anonymous referee for their constructive comments.  Astronomy research at the Open University is funded by a STFC rolling grant.

\bibliographystyle{aa}
\bibliography{mnrasm31}

\appendix

\section[]{Source data}

In this section, we present positional information and spectral fits  in Tables~\ref{pos} and \ref{specdat} for the 185 sources. Table~\ref{pos} provides the right ascension and declination of each source in 2000 coordinates;  these coordinate are not astrometrically corrected, but were taken directly from the source detection routine. We next indicate whether the source is in the D$_{25}$ isophot of NGC\thinspace 253,  the source extraction radius,  background to source area ratio and the encircled energy fraction. Table~\ref{specdat} provides spectral information for each source. We give the best fit emission model for sources with $>$50 source counts in either their pn or MOS spectrum: power  law (PO), blackbody (BB) or bremsstrahlung (BR); if the source is too faint for individual fitting, we show the faint source model applies ( OS or IS, see Table~\ref{specs}).  If an emission model was fitted, then we give the line-of-sight absorption, $n_{\rm H}^{22}$, normalised to an equivalent density of 10$^{22}$ H atom cm$^{-2}$; if the $n_{\rm h}$ falls below the Galactic absorption measured by Stark et al. (1992), 0.013$\times$10$^{22}$,  then the absorption is fixed at this value, indicated by 'f'. The best fit parameter is next, either $\Gamma$ for power law models or kT (in keV) for blackbody or bremsstrahlung models. The best fit $\chi^2$/dof is shown next, with the good fit probability given in square brackets; we only accept models where the good fit probability $>$0.05. Finally, we give the best fit luminosity and the luminosity assuming a standard model. Some models have  the minimum number of counts per bin quoted in parentheses, e.g. PO(30). This indicates that the spectrum had non-standard grouping. Such grouping was done if the standard grouping for a spectrum resulted in a data point that was unusually high or low as an artefact of grouping; grouping the spectrum to e.g. a minimum of 30 counts per bin rather than 50 counts per bin removes the artefact, showing that this outlier is not intrinsic to the source.

\begin{table*}
\caption{Positions of X-ray sources in the XMM-Newton observations of NGC\thinspace 253, along with some observed properties. Column 4 indicates whether the source is within the D$_{25}$ region of NGC\thinspace 253. In Columns 5 and 6, we present the source extraction radius and background to source area ratio respectively. The encircled energy fraction is given for each source region in Column 7..}\label{pos}
\begin{tabular}{ccccccccc}
\noalign{\smallskip}
\hline
\noalign{\smallskip}
{\bf Source} & {\bf $\alpha$} & {\bf $\delta$} & {\bf In D$_{25}$?} & {\bf R$_{\rm s}$} & {\bf A$_{\rm b}$ / A$_{\rm s}$} & {\bf EEF}\\
\noalign{\smallskip}
\hline
\noalign{\smallskip}
1 & 0 46 27.12 & $-$25 21 21.6 & n & 20 & 5.55 & 0.72 \\
2 & 0 46 27.74 & $-$25 19 26.76 & n & 20 & 5.55 & 0.75 \\
3 & 0 46 36.22 & $-$25 17 10.32 & n & 20 & 5.55 & 0.75 \\
4 & 0 46 36.24 & $-$25 20 23.64 & n & 20 & 5.55 & 0.75 \\
5 & 0 46 38.83 & $-$25 12 28.44 & n & 20 & 0 & 0.00 \\
6 & 0 46 43.56 & $-$25 19 35.04 & n & 20 & 5.55 & 0.75 \\
7 & 0 46 44.64 & $-$25 22 35.76 & n & 20 & 4.33 & 0.76 \\
8 & 0 46 46.39 & $-$25 25 53.4 & n & 20 & 12.17 & 0.75 \\
9 & 0 46 47.4 & $-$25 21 50.4 & n & 20 & 4.33 & 0.76 \\
10 & 0 46 48.5 & $-$25 11 47.4 & n & 20 & 10.18 & 0.76 \\
11 & 0 46 49.01 & $-$25 12 29.16 & n & 20 & 10.18 & 0.76 \\
12 & 0 46 51.17 & $-$25 27 30.24 & n & 20 & 10.18 & 0.75 \\
13 & 0 46 55.99 & $-$25 9 42.48 & n & 20 & 10.18 & 0.00 \\
14 & 0 46 56.47 & $-$25 20 34.8 & y & 20 & 5.02 & 0.77 \\
15 & 0 46 56.54 & $-$25 12 51.12 & n & 20 & 10.18 & 0.76 \\
16 & 0 46 56.64 & $-$25 28 39.36 & n & 20 & 10.18 & 0.75 \\
17 & 0 46 56.86 & $-$25 19 1.56 & n & 20 & 5.02 & 0.77 \\
18 & 0 46 57.1 & $-$25 30 26.28 & n & 20 & 5.02 & 0.75 \\
19 & 0 46 57.43 & $-$25 17 43.8 & n & 20 & 2.03 & 0.77 \\
20 & 0 46 59.42 & $-$25 28 46.2 & n & 20 & 10.18 & 0.75 \\
21 & 0 46 59.57 & $-$25 22 54.12 & y & 20 & 12.17 & 0.76 \\
22 & 0 47 0.72 & $-$25 18 33.84 & n & 20 & 5.02 & 0.77 \\
23 & 0 47 0.96 & $-$25 11 16.44 & n & 20 & 10.18 & 0.70 \\
24 & 0 47 1.32 & $-$25 23 25.44 & y & 20 & 12.17 & 0.76 \\
25 & 0 47 1.99 & $-$25 29 29.04 & n & 20 & 10.18 & 0.75 \\
26 & 0 47 2.16 & $-$25 24 24.84 & y & 20 & 12.17 & 0.76 \\
27 & 0 47 3.53 & $-$25 28 51.96 & n & 20 & 10.18 & 0.75 \\
28 & 0 47 5.04 & $-$25 29 36.24 & n & 20 & 10.18 & 0.75 \\
29 & 0 47 5.38 & $-$25 19 42.96 & y & 20 & 5.02 & 0.77 \\
30 & 0 47 5.98 & $-$25 22 27.12 & y & 20 & 5.02 & 0.77 \\
31 & 0 47 6.62 & $-$25 21 27.72 & y & 15 & 8.92 & 0.70 \\
32 & 0 47 6.82 & $-$25 9 11.16 & n & 20 & 5.02 & 0.76 \\
33 & 0 47 6.98 & $-$25 32 22.92 & n & 40 & 1.25 & 0.86 \\
34 & 0 47 7.06 & $-$25 15 48.24 & n & 20 & 2.36 & 0.77 \\
35 & 0 47 8.14 & $-$25 16 24.96 & n & 20 & 8.69 & 0.77 \\
36 & 0 47 9.07 & $-$25 17 39.12 & y & 20 & 2.64 & 0.77 \\
37 & 0 47 9.1 & $-$25 21 24.12 & y & 20 & 5.02 & 0.77 \\
38 & 0 47 9.5 & $-$25 14 4.56 & n & 20 & 12.53 & 0.77 \\
39 & 0 47 10.22 & $-$25 22 33.96 & y & 20 & 5.02 & 0.77 \\
40 & 0 47 10.63 & $-$25 16 31.08 & n & 20 & 8.69 & 0.78 \\
41 & 0 47 11.09 & $-$25 23 33 & y & 20 & 5.06 & 0.62 \\
42 & 0 47 11.57 & $-$25 10 41.16 & n & 20 & 5.02 & 0.77 \\
43 & 0 47 11.98 & $-$25 20 39.12 & y & 20 & 8.69 & 0.78 \\
44 & 0 47 11.98 & $-$25 17 43.44 & y & 20 & 1 & 0.77 \\
45 & 0 47 14.14 & $-$25 29 59.28 & n & 20 & 1.47 & 0.75 \\
46 & 0 47 15.41 & $-$25 12 0.36 & n & 20 & 4 & 0.77 \\
47 & 0 47 15.74 & $-$25 21 46.44 & y & 20 & 1 & 0.77 \\
48 & 0 47 16.1 & $-$25 23 38.76 & y & 20 & 5.06 & 0.77 \\
49 & 0 47 16.9 & $-$25 30 55.44 & n & 20 & 14.06 & 0.75 \\
50 & 0 47 17.5 & $-$25 18 10.43 & y & 10 & 24.5 & 0.58 \\
51 & 0 47 17.52 & $-$25 18 31.47 & y & 10 & 34.75 & 0.58 \\
52 & 0 47 17.83 & $-$25 25 26.76 & n & 20 & 5.06 & 0.77 \\
53 & 0 47 18.26 & $-$25 10 6.6 & n & 20 & 5.02 & 0.77 \\
54 & 0 47 18.46 & $-$25 19 14.88 & y & 20 & 4 & 0.78 \\
55 & 0 47 18.78 & $-$25 21 15.25 & y & 20 & 4 & 0.78 \\
56 & 0 47 19.73 & $-$25 6 45 & n & 20 & 6.68 & 0.76 \\
57 & 0 47 20.14 & $-$25 23 31.56 & y & 20 & 2.6 & 0.57 \\
58 & 0 47 20.33 & $-$25 13 18.48 & n & 20 & 3.52 & 0.78 \\
59 & 0 47 20.45 & $-$25 25 44.76 & n & 20 & 5.06 & 0.77 \\
60 & 0 47 20.88 & $-$25 17 48.48 & y & 20 & 8.69 & 0.78 \\
\noalign{\smallskip}
\hline
\noalign{\smallskip}
\end{tabular}
\end{table*}

\setcounter{table}{0}
\begin{table*}
\caption{continued}
\begin{tabular}{ccccccccc}
\noalign{\smallskip}
\hline
\noalign{\smallskip}
{\bf Source} & {\bf $\alpha$} & {\bf $\delta$} & {\bf In D$_{25}$?} & {\bf R$_{\rm s}$} & {\bf A$_{\rm b}$ / A$_{\rm s}$} & {\bf EEF}\\
\noalign{\smallskip}
\hline
\noalign{\smallskip}
61 & 0 47 20.98 & $-$25 10 1.2 & n & 20 & 2.48 & 0.77 \\
62 & 0 47 22.15 & $-$25 19 35.76 & y & 15 & 3.42 & 0.71 \\
63 & 0 47 22.37 & $-$25 12 1.44 & n & 20 & 2.48 & 0.78 \\
64 & 0 47 22.56 & $-$25 20 51.36 & y & 20 & 4 & 0.78 \\
65 & 0 47 23.09 & $-$25 10 55.2 & n & 20 & 4 & 0.77 \\
66 & 0 47 23.26 & $-$25 30 8.64 & n & 20 & 5.78 & 0.76 \\
67 & 0 47 23.4 & $-$25 19 5.88 & y & 15 & 3.42 & 0.64 \\
68 & 0 47 23.76 & $-$25 15 55.8 & y & 20 & 2.65 & 0.78 \\
69 & 0 47 24.31 & $-$25 14 50.64 & y & 20 & 2.65 & 0.78 \\
70 & 0 47 24.96 & $-$25 18 33.51 & y & 20 & 1.93 & 0.78 \\
71 & 0 47 25.06 & $-$25 19 47.28 & y & 20 & 1.93 & 0.54 \\
72 & 0 47 25.08 & $-$25 21 23.76 & y & 20 & 7.47 & 0.76 \\
73 & 0 47 25.32 & $-$25 16 43.68 & y & 20 & 3.15 & 0.78 \\
74 & 0 47 25.42 & $-$25 5 8.16 & n & 20 & 6.68 & 0.72 \\
75 & 0 47 25.46 & $-$25 28 48.72 & n & 20 & 5.78 & 0.76 \\
76 & 0 47 25.94 & $-$25 20 31.92 & y & 20 & 3.52 & 0.76 \\
77 & 0 47 25.97 & $-$25 8 16.8 & n & 20 & 6.68 & 0.75 \\
78 & 0 47 26.06 & $-$25 33 46.44 & n & 40 & 2.85 & 0.85 \\
79 & 0 47 26.28 & $-$25 15 56.16 & y & 20 & 4 & 0.78 \\
80 & 0 47 26.4 & $-$25 19 14.52 & y & 15 & 3.42 & 0.71 \\
81 & 0 47 26.71 & $-$25 7 38.28 & n & 20 & 6.68 & 0.76 \\
82 & 0 47 27.43 & $-$25 31 9.12 & n & 20 & 5.78 & 0.75 \\
83 & 0 47 27.58 & $-$25 12 20.88 & n & 20 & 2.48 & 0.78 \\
84 & 0 47 27.77 & $-$25 26 44.52 & n & 20 & 4 & 0.74 \\
85 & 0 47 27.88 & $-$25 18 17.05 & y & 20 & 2.1 & 0.66 \\
86 & 0 47 28.22 & $-$25 11 42 & n & 20 & 1.37 & 0.78 \\
87 & 0 47 28.44 & $-$25 9 24.48 & n & 20 & 2.11 & 0.77 \\
88 & 0 47 28.56 & $-$25 10 5.53 & n & 20 & 2.11 & 0.77 \\
89 & 0 47 28.72 & $-$25 19 23.52 & y & 15 & 2.08 & 0.71 \\
90 & 0 47 28.82 & $-$25 16 45.47 & y & 15 & 5.6 & 0.71 \\
91 & 0 47 28.99 & $-$25 28 10.2 & n & 20 & 5.78 & 0.76 \\
92 & 0 47 29.76 & $-$25 21 21.6 & y & 20 & 7.47 & 0.78 \\
93 & 0 47 30.07 & $-$25 17 1.32 & y & 15 & 5.6 & 0.71 \\
94 & 0 47 30.24 & $-$25 8 45.96 & n & 20 & 2.11 & 0.76 \\
95 & 0 47 30.5 & $-$25 11 29.4 & n & 20 & 1.55 & 0.56 \\
96 & 0 47 30.74 & $-$25 18 57.24 & y & 15 & 5.6 & 0.69 \\
97 & 0 47 30.96 & $-$25 18 27.36 & y & 15 & 5.6 & 0.71 \\
98 & 0 47 31.25 & $-$25 15 3.96 & y & 20 & 3.15 & 0.78 \\
99 & 0 47 31.49 & $-$25 10 1.2 & n & 20 & 2.2 & 0.76 \\
100 & 0 47 32.4 & $-$25 28 10.2 & n & 20 & 5.78 & 0.74 \\
101 & 0 47 32.45 & $-$25 31 14.88 & n & 20 & 5.78 & 0.75 \\
102 & 0 47 32.98 & $-$25 17 49.92 & y & 12 & 4.05 & 0.64 \\
103 & 0 47 33 & $-$25 18 45.72 & y & 15 & 5.6 & 0.71 \\
104 & 0 47 33.31 & $-$25 17 22.2 & n & 20 & 5.78 & 0.76 \\
105 & 0 47 33.31 & $-$25 30 19.44 & y & 12 & 8.75 & 0.64 \\
106 & 0 47 33.48 & $-$25 9 9.72 & n & 20 & 2.11 & 0.77 \\
107 & 0 47 33.57 & $-$25 16 33.93 & y & 20 & 3.15 & 0.79 \\
108 & 0 47 35.04 & $-$25 19 13.44 & y & 20 & 3.15 & 0.74 \\
109 & 0 47 35.11 & $-$25 15 12.24 & y & 20 & 3.15 & 0.79 \\
110 & 0 47 35.45 & $-$25 9 35.28 & n & 20 & 2.11 & 0.77 \\
111 & 0 47 35.71 & $-$25 16 31.8 & y & 20 & 3.15 & 0.79 \\
112 & 0 47 36.1 & $-$25 23 58.2 & n & 20 & 6.89 & 0.78 \\
113 & 0 47 37.01 & $-$25 10 42.24 & n & 20 & 2.36 & 0.77 \\
114 & 0 47 37.08 & $-$25 20 3.48 & y & 20 & 8.85 & 0.78 \\
115 & 0 47 38.18 & $-$25 15 43.2 & y & 20 & 3.07 & 0.79 \\
116 & 0 47 38.98 & $-$25 15 2.16 & y & 20 & 3.07 & 0.79 \\
117 & 0 47 39.43 & $-$25 16 32.88 & y & 20 & 3.07 & 0.79 \\
118 & 0 47 40.15 & $-$25 25 32.88 & n & 20 & 5.7 & 0.77 \\
119 & 0 47 40.32 & $-$25 17 57.84 & y & 20 & 1.03 & 0.79 \\
120 & 0 47 40.73 & $-$25 14 11.4 & y & 20 & 5.97 & 0.67 \\

\noalign{\smallskip}
\hline
\noalign{\smallskip}
\end{tabular}
\end{table*}

\setcounter{table}{0}
\begin{table*}
\caption{continued}
\begin{tabular}{ccccccccc}
\noalign{\smallskip}
\hline
\noalign{\smallskip}
{\bf Source} & {\bf $\alpha$} & {\bf $\delta$} & {\bf In D$_{25}$?} & {\bf R$_{\rm s}$} & {\bf A$_{\rm b}$ / A$_{\rm s}$} & {\bf EEF}\\
\noalign{\smallskip}
\hline
\noalign{\smallskip}
121 & 0 47 41.33 & $-$25 16 4.44 & y & 20 & 3.07 & 0.79 \\
122 & 0 47 41.95 & $-$25 17 20.04 & y & 20 & 3.07 & 0.79 \\
123 & 0 47 42.51 & $-$25 14 58.97 & y & 15 & 10.61 & 0.37 \\
124 & 0 47 42.94 & $-$25 9 56.52 & n & 20 & 1.56 & 0.77 \\
125 & 0 47 42.96 & $-$25 13 22.08 & y & 20 & 5.97 & 0.74 \\
126 & 0 47 43.01 & $-$25 15 30.25 & y & 15 & 2.78 & 0.29 \\
127 & 0 47 43.3 & $-$25 6 46.44 & n & 20 & 9 & 0.77 \\
128 & 0 47 44.42 & $-$25 26 51 & n & 20 & 5.7 & 0.77 \\
129 & 0 47 44.64 & $-$25 20 45.6 & n & 20 & 8.85 & 0.78 \\
130 & 0 47 44.86 & $-$25 14 55.3 & y & 15 & 10.61 & 0.71 \\
131 & 0 47 45.05 & $-$25 16 45.12 & y & 20 & 3.07 & 0.78 \\
132 & 0 47 45.1 & $-$25 12 21.24 & y & 20 & 5.97 & 0.78 \\
133 & 0 47 45.79 & $-$25 22 31.8 & n & 20 & 9.55 & 0.78 \\
134 & 0 47 46.66 & $-$25 27 35.64 & n & 20 & 3.07 & 0.76 \\
135 & 0 47 46.73 & $-$25 7 35.73 & n & 20 & 9 & 0.77 \\
136 & 0 47 46.86 & $-$25 29 55.68 & n & 40 & 1.43 & 0.81 \\
137 & 0 47 46.92 & $-$25 8 13.92 & n & 20 & 9 & 0.77 \\
138 & 0 47 47.21 & $-$25 6 30.96 & n & 20 & 9 & 0.76 \\
139 & 0 47 48.26 & $-$25 15 3.96 & y & 20 & 5.97 & 0.76 \\
140 & 0 47 48.34 & $-$25 9 5.04 & n & 20 & 9 & 0.77 \\
141 & 0 47 48.7 & $-$25 12 50.4 & y & 20 & 5.97 & 0.78 \\
142 & 0 47 48.86 & $-$25 16 28.92 & y & 20 & 0 & 0.77 \\
143 & 0 47 49.06 & $-$25 18 10.08 & y & 20 & 8.85 & 0.78 \\
144 & 0 47 49.27 & $-$25 23 8.52 & n & 20 & 9.55 & 0.78 \\
145 & 0 47 49.32 & $-$25 13 35.76 & y & 20 & 5.97 & 0.78 \\
146 & 0 47 50.11 & $-$25 6 59.04 & n & 20 & 9 & 0.77 \\
147 & 0 47 50.28 & $-$25 8 41.64 & n & 20 & 9 & 0.77 \\
148 & 0 47 51.31 & $-$25 10 23.16 & n & 20 & 2.93 & 0.58 \\
149 & 0 47 51.6 & $-$25 24 16.2 & n & 20 & 4.73 & 0.77 \\
150 & 0 47 52.06 & $-$25 17 32.28 & y & 20 & 9.77 & 0.78 \\
151 & 0 47 52.85 & $-$25 7 33.96 & n & 20 & 9 & 0.76 \\
152 & 0 47 53.5 & $-$25 13 8.04 & y & 20 & 3.02 & 0.78 \\
153 & 0 47 55.15 & $-$25 31 1.92 & n & 20 & 5.7 & 0.75 \\
154 & 0 47 55.35 & $-$25 19 6.82 & n & 20 & 9.77 & 0.66 \\
155 & 0 47 56.4 & $-$25 16 18.84 & y & 20 & 9.77 & 0.78 \\
156 & 0 47 57.1 & $-$25 15 3.24 & y & 20 & 0 & 0.78 \\
157 & 0 47 57.22 & $-$25 15 46.41 & y & 20 & 9.77 & 0.54 \\
158 & 0 47 58.01 & $-$25 29 48.48 & n & 20 & 5.7 & 0.76 \\
159 & 0 47 58.22 & $-$25 12 16.92 & n & 20 & 4.73 & 0.77 \\
160 & 0 47 58.25 & $-$25 26 4.2 & y & 20 & 2.93 & 0.75 \\
161 & 0 47 58.73 & $-$25 9 19.44 & n & 20 & 10.99 & 0.76 \\
162 & 0 47 58.92 & $-$25 30 55.44 & n & 20 & 5.7 & 0.75 \\
163 & 0 48 0.07 & $-$25 9 53.64 & y & 20 & 10.99 & 0.77 \\
164 & 0 48 1.01 & $-$25 23 46.65 & n & 20 & 5.1 & 0.77 \\
165 & 0 48 1.22 & $-$25 24 26.31 & n & 20 & 5.1 & 0.72 \\
166 & 0 48 1.27 & $-$25 27 37.44 & n & 20 & 4.73 & 0.76 \\
167 & 0 48 2.09 & $-$25 15 6.84 & y & 20 & 1 & 0.73 \\
168 & 0 48 2.38 & $-$25 11 25.8 & y & 20 & 10.99 & 0.77 \\
169 & 0 48 3.79 & $-$25 12 12.96 & y & 20 & 10.99 & 0.77 \\
170 & 0 48 5.76 & $-$25 25 51.6 & n & 20 & 5.1 & 0.7 \\
171 & 0 48 6.6 & $-$25 12 45 & y & 20 & 10.99 & 0.77 \\
172 & 0 48 7.82 & $-$25 14 17.52 & n & 20 & 5.1 & 0.77 \\
173 & 0 48 7.87 & $-$25 25 7.32 & y & 20 & 7.59 & 0.77 \\
174 & 0 48 8.28 & $-$25 22 57.36 & n & 20 & 11.78 & 0.73 \\
175 & 0 48 9.26 & $-$25 29 42.36 & n & 20 & 9 & 0.69 \\
176 & 0 48 9.46 & $-$25 29 3.84 & n & 20 & 9 & 0.7 \\
177 & 0 48 14.09 & $-$25 19 8.92 & n & 20 & 2.56 & 0.77 \\
178 & 0 48 14.52 & $-$25 13 20.64 & y & 20 & 7.59 & 0.77 \\
179 & 0 48 18.48 & $-$25 13 15.96 & n & 20 & 7.59 & 0.77 \\
180 & 0 48 18.84 & $-$25 15 8.28 & n & 20 & 7.59 & 0.77 \\

\noalign{\smallskip}
\hline
\noalign{\smallskip}
\end{tabular}
\end{table*}

\setcounter{table}{0}
\begin{table*}
\caption{continued}
\begin{tabular}{ccccccccc}
\noalign{\smallskip}
\hline
\noalign{\smallskip}
{\bf Source} & {\bf $\alpha$} & {\bf $\delta$} & {\bf In D$_{25}$?} & {\bf R$_{\rm s}$} & {\bf A$_{\rm b}$ / A$_{\rm s}$} & {\bf EEF}\\
\noalign{\smallskip}
\hline
\noalign{\smallskip}
181 & 0 48 18.98 & $-$25 14 21.48 & n & 20 & 7.59 & 0.77 \\
182 & 0 48 23.04 & $-$25 19 12.36 & n & 20 & 11.1 & 0.77 \\
183 & 0 48 24.53 & $-$25 18 11.16 & n & 20 & 11.1 & 0.76 \\
184 & 0 48 31.25 & $-$25 23 51 & n & 20 & 11.78 & 0.76 \\
185 & 0 48 33.36 & $-$25 15 19.25 & n & 40 & 1.18 & 0.11 \\

\noalign{\smallskip}
\hline
\noalign{\smallskip}
\end{tabular}
\end{table*}

\begin{table*}
\setlength{\tabcolsep}{2.5pt}
 \caption{ Spectral properties for each source. For each source we show the net counts in the pn and MOS detectors. 
We show the best fit model, absorption, parameter (spectral index or temperature), $\chi^2$/dof and good fit probability, best fit luminosity and standard model  (Method I) luminosity. Best fit models for bright sources can be a power law (PO), blackbody (BB), bremsstrahlung (BR), or a two component model consisting of a blackbody plus power law (2C). Faint sources are modelled using a best fit power law listed in Table~\ref{specs}. Numbers in parentheses represent uncertainties in the last digit at the 90\% confidence level. The quality of the spectrum may be deduced from the number of degrees of freedom because the spectra are grouped to a minimum number of counts per bin (brighter sources have more degrees of freedom). }\label{specdat}
\begin{tabular}{cccccccccccccccc}
\noalign{\smallskip}
\hline
\noalign{\smallskip}
\footnotesize

{\bf S}  & {\bf pn source counts}& {\bf MOS source counts}& {\bf Mod} & {\bf $n_{\rm H}^{22}$}& {\bf Par} & {\bf $\chi^2$/dof [gf] }& {\bf $L_{36}^{\rm BF}$} & {\bf $L_{36}^{SM}$ }\\
\noalign{\smallskip}
\hline
\noalign{\smallskip}
1 & 48 & 0 &   OS   &      &     &    &  11.6(13)  &  8.04 \\
2 & 27 & 0 &   OS   &      &     &     &  6.3(7)  &  4.35 \\
3 & 0 & 0 &  X  &    &    &     &     &  0 \\
4 & 34 & 3 &   OS   &      &     &     &  7.8(8)  &  5.43 \\
5 & 0 & 0 &  X  &      &     &     &     &  0 \\
6 & 57 & 68 &   PO   &   0.09(9)   &   2.5(12)   &   2/8 [0.92]   &   27(12)   &  9.16 \\
7 & 53 & 31 &   PO   &    0.4(4)   &   2.3(1.9)   &   0.5/3 [0.915]   &   46(30)   &  8.43 \\
8 & 57 & 30 &   PO   &    f   &   1.9(13)   &   6/4 [0.18]   &   29(27)   &  9.11 \\
9 & 709 & 596 &   PO   &    f   &   1.9(9)   &   18/23 [0.78]   &   240(33)   &  112.64 \\
10 & 33 & 76 &   PO   &    f   &   1.3(5)   &   5/5 [0.46]   &   46(30)   &  5.23 \\
11 & 41 & 60 &   PO   &    f   &   2.1(7)   &   0.5/3 [0.915]   &   50(25)   &  6.48 \\
12 & 0 & 0 &  X  &      &     &     &     &  0 \\
13 & 0 & 0 &  X  &      &     &     &     &  0 \\
14 & 164 & 129 &   PO   &   0.13(11)   &   1.8(4)   &   38/43 [0.67]   &   67(19)   &  25.73 \\
15 & 359 & 306 &   PO   &   0.10(5)   &   1.83(17)   &   39/37 [0.39]   &   150(30)   &  56.39 \\
16 & 0 & 0 &  X  &      &     &     &     &  0 \\
17 & 53 & 28 &   PO   &    f   &   0.2(1.4)   &   8/6 [0.27]   &   40(30)   &  8.3 \\
18 & 96 & 0 &   PO   &    f   &   1.7(6)   &   12/15 [0.69]   &   61(35)   &  15.47 \\
19 & 154 & 124 &   PO   &   0.24(13)   &   2.6(6)   &   46/38 [0.18]   &   51(15)   &  24.09 \\
20 & 0 & 21 &   OS   &      &     &     &  5.9(6)  &  4.09 \\
21 & 109 & 61 &   PO   &    f   &   1.0(3)   &   32/22 [0.07]   &   52(29)   &  17.13 \\
22 & 36 & 33 &   OS   &      &     &     &  8.1(9)  &  5.62 \\
23 & 49 & 52 &   PO   &    f   &   2.7(12)   &   3 / 4 [0.63]   &   19(13)   &  8.4 \\
24 & 118 & 98 &   PO   &    f   &   1.6(3)   &   35/34 [0.44]   &   36(17)   &  18.54 \\
25 & 5 & 0 &   OS   &      &     &     &  1.15(13)  &  0.8 \\
26 & 67 & 58 &   PO   &    f   &   2.2(5)   &   5/10 [0.88]   &   19(12)   &  10.55 \\
27 & 30 & 16 &   OS   &      &     &     &  6.9(8)  &  4.78 \\
28 & 10 & 0 &   OS   &      &     &     &  2.3(3)  &  1.6 \\
29 & 1034 & 930 &   PO   &   0.17(3)   &   2.9(2)   &   41/35 [0.24]   &   355(38)   &  161.92 \\
30 & 102 & 79 &   BB   &   0.7(3)   &   0.08(2)   &   8/6 [0.25]   &   2976(2972)   &  15.94 \\
31 & 96 & 112 &   PO   &    f   &   1.2(2)   &   29/28 [0.42]   &   38(19)   &  16.46 \\
32 & 119 & 0 &   PO   &   0.3(2)   &   2.3(8)   &   8/10 [0.58]   &   52(19)   &  18.71 \\
33 & 181 & 0 &   PO   &    f   &   1.8(4)   &   12/24 [0.98]   &   90(46)   &  25.36 \\
34 & 118 & 65 &   PO   &   0.12(11)   &   2.5(11)   &   16/20 [0.73]   &   27(12)   &  18.33 \\
35 & 326 & 184 &   PO   &   0.04(4)   &   2.3(4)   &   35/40 [0.70]   &   61(12)   &  50.58 \\
36 & 51 & 42 &   PO   &   0.4(4)   &   0.10(6)   &   1/6 [0.99]   &   42(42)   &  7.91 \\
37 & 3244 & 3464 &   PO   &   0.53(10)   &   0.71(10) 2.0(2)   &   117/117 [0.48]   &   1762(135)   &  504.88 \\
38 & 524 & 323 &   PO   &    f   &   2.30(12)   &   35/27 [0.15]   &   84(15)   &  81.38 \\
39 & 192 & 161 &   PO   &   0.13(13)   &   0.8(6)   &   50/50[0.49]   &   134(38)   &  29.93 \\
40 & 28 & 14 &   OS   &      &     &     &  6.3(7)  &  4.34 \\
41 & 434 & 489 &   PO   &   0.18(4)   &   1.66(16)   &   34/47 [0.92]   &   388(25)   &  84.01 \\
42 & 130 & 0 &   PO   &   0.19(16)   &   2.2(6)   &   13/10 [0.20]   &   42(15)   &  20.3 \\
43 & 34 & 38 &   IS   &      &     &     &  14(3)  &  5.26 \\
44 & 221 & 174 &   BB   &   0.5(2)   &   0.12(3)   &   45/40 [0.26]   &   202(83)   &  34.52 \\
45 & 0 & 7 &   OS   &      &     &     &  1.89(19)  &  1.39 \\
46 & 184 & 24 &   PO   &   0.08(8)   &   1.8(5)   &   34/31 [0.34]   &   44(13)   &  28.58 \\
47 & 267 & 199 &   PO   &   0.07(7)   &   1.5(3)   &   33/45 [0.90]   &   88(21)   &  41.62 \\
48 & 26 & 22 &   IS   &      &     &     &  11(2)  &  4.05 \\
49 & 50 & 0 &   PO   &    f   &   0.13(5)   &   4/4 [0.44]   &   12(10)   &  8 \\
50 & 3924 & 4144 &   PO   &   0.38(8)   &   2.10(8)   &   158/133 [0.07]   &    1890(150)   &  816.53 \\
51 & 882 & 814 &   PO   &   0.43(6)   &   2.48(17)   &   30/28 [0.36]   &   516(65)   &  183.53 \\
52 & 65 & 54 &   PO   &   0.5(4)   &   2.7(12)   &   8/12 [0.75]   &   27(12)   &  10.17 \\
53 & 154 & 177 &   PO   &   0.30(18)   &   1.4(4)   &   38/37 [0.45]   &   83(25)   &  24 \\
54 & 1001 & 1002 &   PO     &   0.12(3)   &   1.51(8)   &   46/42 [0.32]   &   297(34)   &  154.34 \\
55 & 0 & 0 &  X  &      &     &     &     &  0 \\
56 & 171 & 167 &   BB   &    f   &   0.193(16)   &   49/58   &   25(8)   &  26.93 \\
57 & 31 & 91 &   PO   &   0.6(5)   &   2.8(10)   &   0.4/6 [0.999]   &   56(21)   &  6.55 \\

\noalign{\smallskip}
\hline
\noalign{\smallskip}
\end{tabular}
\end{table*}

\setcounter{table}{1}
\begin{table*}
\setlength{\tabcolsep}{2pt}
\caption{continued}
\begin{tabular}{cccccccccccccccc}
\noalign{\smallskip}
\hline
\noalign{\smallskip}
\footnotesize

{\bf S}  & {\bf pn source counts}& {\bf MOS source counts}& {\bf Mod} & {\bf $n_{\rm H}^{22}$}& {\bf Par} & {\bf $\chi^2$/dof [gf] }& {\bf $L_{36}^{\rm BF}$} & {\bf $L_{36}^{SM}$ }\\
\noalign{\smallskip}
\hline
\noalign{\smallskip}
58 & 235 & 71 &   BB   &    f   &   0.20(2)   &   40/28 [0.07]   &   19(6)   &  36.3 \\
59 & 145 & 81 &   PO   &   0.04(4)   &   2.8(8)   &   33/29 [0.26]   &   27(10)   &  22.69 \\
60 & 216 & 256 &   PO   &   0.23(13)   &   1.6(3)   &   45/36 [0.14]   &   77(17)   &  33.21 \\
61 & 68 & 58 &   PO   &    f   &   1.6(9)   &   20/18 [0.33]   &   19(15)   &  10.59 \\
62 & 202 & 132 &   PO   &    f   &   2.3(4)   &   47/34 [0.07]   &   27(13)   &  34.28 \\
63 & 93 & 140 &   PO   &    4(3)   &   1.6(9)   &   15/23 [0.89]   &   92(46)   &  14.39 \\
64 & 10537 & 10390 &   2C   &   0.20(2)   &   0.73(5) 2.14(4)   &   347/323 [0.17]   &   2911(119)   &  1625.16 \\
65 & 577 & 469 &   PO   &   0.05(4)   &   1.78(14)   &   33/43 [0.87]   &   138(23)   &  89.55 \\
66 & 0 & 0 &   X   &      &     &     &     &  0 \\
67 & 1390 & 1670 &   2C   &   0.11(3)   &   0.78(13) 2.1(2)   &   68/57 [0.14]   &   490(54)   &  262.39 \\
68 & 112 & 95 &   PO   &   0.05(5)   &   2.0(5)   &   22/23 [0.54]   &   19(8)   &  17.19 \\
69 & 257 & 87 &   PO   &   0.14(11)   &   2.0(4)   &   31/29 [0.35]   &   56(13)   &  39.5 \\
70 & 658 & 439 &   BB   &   0.17(13)   &   0.15(5)   &   38/26 [0.27]   &   50(38)   &  100.96 \\
71 & 441 & 781 &   2C   &   0.2(2)   &   1.3(3) 3.3(1.5)   &   63/46 [0.05]   &   290(69)   &  97.51 \\
72 & 218 & 245 &   PO   &   0.04(4)   &   1.7(3)   &   39/29 [0.10]   &   48(12)   &  34.28 \\
73 & 213 & 227 &   PO   &   0.48(17)   &   2.0(3)   &   39/38 [0.44]   &   83(19)   &  32.65 \\
74 & 147 & 69 &   PO   &    f   &   1.6(4)   &   24/22 [0.36]   &   73(27)   &  24.39 \\
75 & 69 & 42 &   PO   &   0.2(2)   &   2.8(19)   &   5/4 [0.32]   &   31(15)   &  10.9 \\
76 & 83 & 109 &   BB   &   0.14(14)   &   0.11(5)   &   19/30 [0.94]   &   27(6)   &  13.07 \\
77 & 112 & 99 &   PO   &    f   &   2.1(4)   &   48/35 [0.07]   &   21(12)   &  17.83 \\
78 & 0 & 0 &  X  &      &     &     &     &  0 \\
79 & 228 & 90 &   BB   &   0.6(3)   &   0.10(3)   &   45/35 [0.12]   &   159(127)   &  34.95 \\
80 & 228 & 90 &   BB   &   0.5(3)   &   0.09(3)   &   45/35 [0.12]   &   346(309)   &  38.6 \\
81 & 94 & 76 &   BR   &   0.14(14)   &   0.45(25)   &   18/20 [0.57]   &   29(21)   &  14.93 \\
82 & 0 & 0 &   X   &      &     &     &     &  0 \\
83 & 106 & 68 &   PO   &    f   &   1.2(4)   &   25/20 [0.21]   &   33(17)   &  16.36 \\
84 & 73 & 76 &   PO   &    f   &   2.0(4)   &   23/18 [0.19]   &   13(10)   &  11.79 \\
85 & 528 & 878 &   PO   &   0.05(5)   &   1.56(17)   &   41/36 [0.28]   &   179(38)   &  96.82 \\
86 & 76 & 40 &   BB   &   0.2(2)   &   0.08(8)   &   7/7 [0.43]   &   15(13)   &  11.76 \\
87 & 65 & 7 &   BB   &    f   &   0.15(5)   &   6/7 [0.52]   &   6.7(5)   &  10.12 \\
88 & 70 & 54 &   PO   &    f   &   2.0(8)   &   17/13 [0.22]   &   23(13)   &  10.95 \\
89 & 350 & 399 &   PO   &   0.9(4)   &   1.7(3)   &   67/51 [0.07]   &   140(29)   &  59.22 \\
90 & 305 & 282 &   PO   &   0.37(16)   &   1.6(3)   &   40/55 [0.93]   &   108(25)   &  51.52 \\
91 & 106 & 75 &   PO   &   0.07(7)   &   3.0(12)   &   17/22 [0.76]   &   29(10)   &  16.69 \\
92 & 98 & 73 &   PO   &    f   &   2.0(7)   &   17/17 [0.43]   &   13(10)   &  15.09 \\
93 & 753 & 695 &   2C   &   0.5(2)   &   0.13(2) 2.1(4)   &   44/32 [0.07]   &   395(124)   &  127.12 \\
94 & 46 & 7 &   OS   &      &     &     &  10.5(11)  &  7.31 \\
95 & 18 & 58 &   PO   &    f   &   0.6(7)   &   5/7 [0.67]   &   36(27)   &  3.84 \\
96 & 369 & 356 &   PO   &   0.10(5)   &   3.6(6)   &   61/56 [0.32]   &   77(15)   &  63.84 \\
97 & 247 & 284 &   PO   &   0.4(4)   &   1.5(6)   &   21/19 [0.34]   &   434(257)   &  41.71 \\
98 & 409 & 453 &   PO   &   0.41(13)   &   1.6(2)   &   75/63 [0.14]   &   142(27)   &  62.62 \\
99 & 188 & 192 &   PO   &   0.04(4)   &   1.9(3)   &   51/54 [0.58]   &   42(12)   &  29.56 \\
100 & 250 & 204 &   BB   &    1.3(3)   &   0.058(9)   &   28/23 [0.20]   &   1.4(14)  E+5   &  33.65 \\
101 & 0 & 0 &   X   &      &     &     &     &  0 \\
102 & 11614 & 12014 &   2C    &   0.29(2)   &   0.98(6) 1.94(5)   &   374/374 [0.49]   &   4064(193)   &  2173.73 \\
103 & 343 & 292 &   BB   &   0.11(8)   &   0.11(2)   &   56/54 [0.39]   &   50(10)   &  57.91 \\
104 & 0 & 0 &   X   &      &     &     &     &  0 \\
105 & 18597 & 18808 &   can't fit   &      &     &     &   8227(785)   &  3480.01 \\
106 & 40 & 2 &   OS   &      &     &     &  9.0(10)  &  6.23 \\
107 & 1476 & 1232 &   PO(20)   &   0.07(4)   &   2.00(15)   &   131/118 [0.20]   &   193(44)   &  225.17 \\
108 & 439 & 504 &   PO   &    f    &   1.97(15)   &   50/57 [0.72]   &   73(19)   &  71.05 \\
109 & 4281 & 4795 &   2C   &   0.66(9)   &   0.77(4) 3.1(2)   &   157/165 [0.65]   &   2357(293)   &  654.38 \\
110 & 31 & 14 &   OS   &      &     &     &  6.9(8)  &  4.82 \\
111 & 1950 & 1775 &   PO   &   0.093(10)   &   1.81(8)   &   108/85   &   359(31)   &  297.15 \\
112 & 24 & 6 &   OS   &      &     &     &  5.4(6)  &  3.72 \\
113 & 13 & 0 &   OS   &      &     &     &  2.9(3)  &  2.02 \\
114 & 22 & 28 &   IS   &      &     &     &  9.3(19)  &  3.37 \\
115 & 762 & 568 &    PO   &   0.32(13)   &   0.15(2)   &   30/20 [0.06]   &   205(38)   &  116.25 \\
116 & 130 & 212 &   PO   &   0.5(2)   &   3.0(5)   &   36/41 [0.67]   &   69(35)   &  19.87 \\
117 & 1029 & 729 &   BB   &   0.25(10)   &   0.132(15)   &   44/31 [0.06]   &   500(135)   &  156.61 \\
118 & 0 & 0 &   X   &      &     &     &     &  0 \\
119 & 383 & 247 &   BB   &   0.12(10)   &   0.15(3)   &   46/34 [0.08]   &   40(10)   &  58.33 \\
120 & 275 & 570 &   PO   &   0.24(4)   &   1.79(19)   &   19/28 [0.90]   &   146(31)   &  49.37 \\

\noalign{\smallskip}
\hline
\noalign{\smallskip}
\end{tabular}
\end{table*}

\setcounter{table}{1}
\begin{table*}
\setlength{\tabcolsep}{3.pt}
\caption{continued}
\begin{tabular}{ccccccccccccccccccc}
\noalign{\smallskip}
\hline
\noalign{\smallskip}
{\bf S} & {\bf pn source counts} & {\bf MOS source counts} &  {\bf Mod} & {\bf $n_{\rm H}^{22}$}& {\bf Par} & {\bf $\chi^2$/dof [gf] }& {\bf $L_{36}^{\rm BF}$} & {\bf $L_{36}^{SM}$ }\\
\noalign{\smallskip}
\hline
\noalign{\smallskip}
121 & 943 & 634 &   BB (20)   &   0.35(11)   &   0.12(2)   &   83/69   &   434(71)   &  143.74 \\
122 & 282 & 137 &   BB      &    f   &   0.16(2)   &   43/45 [0.57]   &   15(4)   &  42.91 \\
123 & 773 & 4156 &   2C    &   0.60(11)   &   0.94(8) 3.4(5)   &   84/76 [0.24]   &   2376(389)   &  249.8 \\
124 & 135 & 100 &   PO   &   0.08(8)   &   1.6(6)   &   32/37 [0.68]   &   36(12)   &  20.95 \\
125 & 314 & 371 &   PO   &   0.24(8)   &   1.60(17)   &   49/42   &   123(23)   &  51.03 \\
126 & 173 & 662 &   PO   &   0.25(8)   &   2.1(2)   &   53/42 [0.12]   &   276(86)   &  71.72 \\
127 & 49 & 76 &   PO   &   0.2(2)   &   3.3(1.9)   &   7/4 [0.16]   &   25(13)   &  7.69 \\
128 & 420 & 313 &   PO   &    f    &   1.82(12)   &   31/40 [0.85]   &   106(21)   &  65.76 \\
129 & 104 & 98 &   PO   &   0.3(2)   &   3.1(11)   &   33/24 [0.10]   &   33(13)   &  15.97 \\
130 & 415 & 332 &   PO   &   0.15(5)   &   1.95(16)   &   38/39 [0.51]   &   100(15)   &  70.12 \\
131 & 337 & 219 &   BB   &   0.4(3)   &   0.08(2)     &   57/47 [0.16]   &   614(403)   &  51.61 \\
132 & 80 & 48 &   PO (30)   &    f    &   1.7(7)   &   10/7 [0.22]   &   19(13)   &  12.33 \\
133 & 72 & 27 &   PO   &    f   &   1.9(9)   &   7/7 [0.39]   &   12(10)   &  11.11 \\
134 & 0 & 43 &   OS   &      &     &     &  11.3(7)  &  8.31 \\
135 & 38 & 62 &   PO    &   0.2(2)   &   2.5(15)   &   3/3 [0/40]   &   19(12)   &  5.95 \\
136 & 0 & 127 &   PO    &   0.02(2)   &   3.1(14)   &   16/19 [0.62]   &   36(17)   &  21.26 \\
137 & 65 & 76 &   PO     &   0.07(7)   &   2.0(9)   &   12/11 [0.35]   &   17(10)   &  10.15 \\
138 & 122 & 155 &   PO     &   0.07(7)   &   1.5(4)   &   35/41 [0.71]   &   56(19)   &  19.17 \\
139 & 384 & 362 &   PO   &   f   &   1.1(2)   &   39/40 [0.52]   &   84(29)   &  60.61 \\
140 & 124 & 132 &   PO    &   0.13(11)   &   1.5(4)   &   27/37 [0.81]   &   60(17)   &  19.45 \\
141 & 127 & 179 &   PO   &   0.7(3)   &   1.5(3)   &   52/40 [0.09]   &   83(23)   &  19.56 \\
142 & 196 & 158 &   BB      &    f   &   0.17(2)   &   41/30 [0.08]   &   13(6)   &  30.73 \\
143 & 117 & 91 &   PO    &   0.13(6)   &   3.2(12)   &   13/24 [0.96]   &   19(8)   &  17.93 \\
144 & 57 & 85 &   PO    &   0.9(8)   &   1.2(7)   &   11/14 [0.73]   &   44(19)   &  8.82 \\
145 & 177 & 146 &   BB      &   0.5(3)   &   0.10(3)   &   26/28 [0.57]   &   274(183)   &  27.21 \\
146 & 52 & 35 &   PO     &    f   &   1.1(9)   &   7/5 [0.25]   &   33(25)   &  8.16 \\
147 & 490 & 328 &   PO    &    f   &   2.02(12)   &   38/43 [0.71]   &   102(19)   &  76.46 \\
148 & 0 & 72 &   PO    &   0.7(7)   &   2.4(19)   &   9/6 [0.20]   &   38(23)   &  14.1 \\
149 & 69 & 83 &   PO    &    1.9(16)   &   3.4(16)   &   15/12 [0.23]   &   52(29)   &  10.72 \\
150 & 101 & 58 &   BB   &   0.3(3)   &   0.11(7)   &  5/6 [0.59]  &   27(8)   &  15.47 \\
151 & 177 & 67 &   PO    &   0.34(10)   &   2.6(4)   &   41/46 [0.70]   &   106(27)   &  27.89 \\
152 & 171 & 166 &   PO    &   0.014(14)   &   1.7(3)   &   37/39 [0.55]   &   33(12)   &  26.41 \\
153 & 0 & 0 &   X   &    &    &     &     &  0 \\
154 & 84 & 61 &   BB   &    f   &   0.10(2)   &   17/11 [0.11]   &   17(10)   &  15.19 \\
155 & 295 & 250 &   PO   &    f   &   1.78(15)   &   27/35 [0.82]   &   50(12)   &  45.39 \\
156 & 27 & 21 &   IS   &      &     &     &  11(2)  &  4.15 \\
157 & 13 & 28 &   IS   &      &     &     &  7.9(15)  &  2.88 \\
158 & 32 & 8 &   OS   &      &     &     &  7.3(8)  &  5.08 \\
159 & 88 & 94 &   PO    &    f   &   2.1(4)   &   15/14 [0.39]   &   21(13)   &  13.78 \\
160 & 109 & 141 &   PO    &   0.16(14)   &   1.9(5)   &   33/31 [0.37]   &   35(12)   &  17.48 \\
161 & 107 & 59 &   PO     &   0.03(3)   &   2.2(7)   &   19/21 [0.60]   &   25(10)   &  16.88 \\
162 & 11 & 1 &   OS   &      &     &     &  2.5(3)  &  1.76 \\
163 & 1887 & 1722 &   PO     &   0.03(2)   &   1.75(8)   &   60/65 [0.66]   &   516(38)   &  294.7 \\
164 & 512 & 428 &   PO    &    f   &   1.97(12)   &   31/31 [0.47]   &   111(21)   &  79.72 \\
165 & 103 & 103 &   PO    &    f   &   1.9(4)   &   34/24 [0.09]   &   25(13)   &  17.1 \\
166 & 79 & 59 &   PO    &   0.05(5)   &   3.0(13)   &   17/12 [0.14]   &   23(10)   &  12.46 \\
167 & 15 & 0 &   IS   &      &     &     &  6.7(13)  &  2.46 \\
168 & 87 & 99 &   PO    &    f   &   1.9(4)   &   13/15 [0.57]   &   17(10)   &  13.52 \\
169 & 80 & 50 &   PO   &    f   &   3.2(8)   &   8/12 [0.78]   &   10(6)   &  12.42 \\
170 & 35 & 41 &   OS   &      &     &     &   4(3)   &  6.02 \\
171 & 118 & 61 &   BB   &    f   &   0.16(2)   &   3/14 [0.99]   &   12(5)   &  18.33 \\
172 & 171 & 137 &   PO   &    f   &   2.1(3)   &   43/47 [0.62]   &   40(13)   &  26.82 \\
173 & 0 & 22 &   IS   &      &     &     &  10.6(3)  &  3.88 \\
174 & 72 & 43 &   PO    &   0.04(4)   &   1.8(9)   &   1.2/3 [0.75]   &   35(17)   &  11.91 \\
175 & 15 & 0 &   OS   &      &     &     &  3.7(4)  &  2.62 \\
176 & 9 & 39 &   OS   &      &     &     &  2.2(3)  &  1.55 \\
177 & 60 & 76 &   PO    &    f   &   1.5(4)   &   8/12 [0.79]   &   21(13)   &  9.34 \\
178 & 0 & 3 &   IS   &      &     &     &  1.4(3)  &  0.59 \\
179 & 161 & 96 &   PO     &   0.05(5)   &   1.8(3)   &   16/30 [0.98]   &   46(13)   &  25.19 \\
180 & 84 & 82 &   PO    &   0.9(8)   &   1.1(8)   &   8/12 [0.77]   &   81(31)   &  13.12 \\

\noalign{\smallskip}
\hline
\noalign{\smallskip}
\end{tabular}
\end{table*}

\setcounter{table}{1}
\begin{table*}
\setlength{\tabcolsep}{5.4pt}
\caption{continued}
\begin{tabular}{ccccccccccccccccccc}
\noalign{\smallskip}
\hline
\noalign{\smallskip}
{\bf S} &{\bf pn source counts}&{\bf MOS source counts}& {\bf Mod} & {\bf $n_{\rm H}^{22}$}& {\bf Par} & {\bf $\chi^2$/dof [gf] }& {\bf $L_{36}^{\rm BF}$} & {\bf $L_{36}^{SM}$ }\\
\noalign{\smallskip}
\hline
\noalign{\smallskip}
181 & 23 & 28 &   OS  &   &   &   &  5.2(6)  &  3.6 \\
182 & 126 & 89 &  PO   &  0.03(3)  &  3.2(9)  &  16/24 [0.87]  &  25(8)  &  19.76 \\
183 & 45 & 41 &   OS  &   &   &   &  10.2(12)  &  7.08 \\
184 & 132 & 38 &  PO   &  0.02(2)  &  2.0(6)  &  18/21 [0.67]  &  60(19)  &  21\\
185 & 0 & 40 &  OS  &   &   &   &   9.6(9)  &  6.71 \\

\noalign{\smallskip}
\hline
\noalign{\smallskip}
\end{tabular}
\end{table*}

\bsp

\label{lastpage}

\end{document}